\documentclass[useAMS,usenatbib,usegraphicx]{mn2e}

\title[Radio imaging Circinus X-1]{A decade of radio imaging the relativistic outflow in the peculiar X-ray binary Circinus X-1} 
\author[V. Tudose et al.]{V. Tudose,$^{1,2,3}$\thanks{E-mail: vtudose@science.uva.nl (VT)}
R.P. Fender,$^{4,1}$ A.K. Tzioumis,$^{5}$ R.E. Spencer$^{6}$ and \newauthor M. van
der Klis$^{1}$ \\ $^{1}$"Anton Pannekoek"
Astronomical Institute, University of Amsterdam, Kruislaan 403, 1098
SJ Amsterdam, the Netherlands\\ $^{2}$Astronomical Institute of the
Romanian Academy, Cutitul de Argint 5, RO-040557 Bucharest, Romania\\
$^{3}$Research Center for Atomic Physics and Astrophysics, Atomistilor 405, RO-077125 Bucharest, Romania \\
$^{4}$School of Physics and Astronomy, University of Southampton,
Highfield, SO17 1BJ Southampton \\ $^{5}$Australia Telescope
National Facility, CSIRO, PO Box 76, Epping New South Wales 1710,
Australia \\ $^{6}$Jodrell Bank Observatory, University of Manchester, 
SK11 9DL Cheshire}

\begin{document}

\date{Accepted 2008 July 29. Received 2008 July 29; in original form 2008 March 28}

\pagerange{\pageref{firstpage}--\pageref{lastpage}} \pubyear{2008}

\maketitle

\label{firstpage}

\begin{abstract}
\
We present observations of the neutron star X-ray binary and relativistic jet source Circinus X-1 made at 4.8 and 8.6 GHz with the Australia Telescope 
Compact Array during a time interval of 
almost 10 years. The system shows significant variations in the morphology and brightness of the radio features on all timescales from days to years.
Using the time delay between the successive brightening of the different components of the radio emission we were able 
to provide further evidence for 
the relativistic nature of the arcsec scale outflow, with an apparent velocity $\beta_{app} \geq$ 12.
No compelling 
evidence for an evolution of the orientation of the jet axis was found.  We also place an upper limit on the proper motion of the system which is 
consistent with previous optical studies. Besides the 
previously reported radio flares close to the orbital phase 0.0 (interpreted as enhanced accretion at periastron passage), we also identified 
outbursts with similar properties near the orbital phase 0.5. The global spectral index revealed a preferentially 
steep spectrum over the entire period of monitoring with a mean value and standard deviation $\alpha$=-0.9 $\pm$ 0.6 ($F_{\nu} \propto \nu^{\alpha}$), 
which became significantly flatter 
during the outbursts. Polarization was detected in one third of the epochs and in one case Faraday rotation close to the core of the system was measured. 
\end{abstract}

\begin{keywords}
accretion, accretion discs -- stars: individual: Circinus X-1 --  ISM: jets and outflows -- radio 
continuum: stars -- radiation mechanisms: non-thermal -- X-rays: stars.
\end{keywords}	

\section{Introduction}

Circinus X-1 is one of the most exotic X-ray binary systems. The detection of type I X-ray bursts \citep*{Ten86a,Ten86b} and twin kHz quasi-periodic 
oscillations in the X-ray power spectra \citep{Bou06} strongly suggests that the compact object in the system is a neutron star. The nature of 
the companion star is more debatable: some evidence is pointing towards a low-mass object \citep{Joh99}, while other towards a high-mass one 
\citep*{Mur80,Jon07}. If the companion star is a supergiant, as recent evidence seems to suggest \citep{Jon07}, then this is at odds with the low magnetic 
field that must be present in the system (as inferred from type I X-ray bursts observations, for instance). Thus Circinus X-1 appears to be a very exotic 
system, harbouring a young neutron star with a low magnetic field. 

Neutron star X-ray binaries (XRBs) are classified according to their X-ray spectral and timing properties as Z and Atoll sources. The Z class comprises seven 
objects accreting near the Eddington limit. All of them have been detected in the radio band and show variable emission at cm wavelengths. The Atoll type 
form the largest class of neutron star XRBs and have X-ray properties similar to black hole XRBs. Due to their systematic lower radio fluxes, one order of 
magnitude less radio loud than Z sources, only five of them have been detected in the radio band. The behaviour of Circinus X-1 is puzzling, exhibiting 
similarities with Z sources (e.g. \citealt*{Shi99}), but 
at times showing characteristics reminiscent of Atoll sources \citep{Oos95}. The orbit of the binary system probably has a relatively high 
eccentricity (e $\simeq$ 0.4-0.9) that produces variations in the accretion rate on the compact object \citep*{Mur80,Nic80,Tau99,Jon07}. This offers an 
explanation for the periodic flares (P$\simeq$16.6 d) observed in X-ray \citep{Kal76}, infrared \citep{Gla78,Gla94} and radio \citep{Whe77,Hay78}, which 
are interpreted as enhanced accretion near the periastron passage. 

Circinus X-1 is associated with an arcmin scale synchrotron nebula \citep{Hay86,Ste93} 
which is probably powered by the jet originating close to the binary system \citep{Hei02,Tud06}. The jet is observed in radio at arcmin as well as arcsec 
scales and was recently detected in X-ray band by Chandra (\citealt{Hei07}; Soleri et al. 2008). Observations at cm wavelengths offered evidence 
for the presence of a relativistic outflow aligned very close to the line of sight \citep{Fen04}. However, X-ray \citep{Sch08,Iar08} and optical \citep{Jon07} 
spectroscopy seem to favour a significantly higher angle between the jet axis and the line of sight. 

Besides the flux modulation at the orbital 
period, Circinus X-1 also exhibits `secular' changes, documented in X-ray and radio. The ASM/RXTE (All Sky Monitor/Rossi X-ray Timing Explorer) light curve 
(2-10 keV) shows that after a quasi-constant X-ray output, the flux decreased starting from 2000 up to now by one order of magnitude (Fig. 1, left, 
bottom panel). The picture is even more dramatic in radio. In the 1970s-1980s the radio flares reached up to 1 Jy (e.g. \citealt{Whe77,Hay78}), then they 
were detected at mJy level only \citep{Ste91,Fen97,Fen98}, until recently when a relatively increased radio activity from Circinus X-1 was noticed 
\citep*{Fen05,Nic07, Del07}. This offered the opportunity to obtain new VLBI (Very Large Baseline Interferometry) observations of the system within the 
framework of the first Southern e-VLBI (electronic VLBI) experiment \citep{Phi07}, after almost 25 years since the last successful VLBI observations \citep{Pre83}.

Circinus X-1 complex is very close to the supernova remnant SNR G321.9-0.3 on the plane of the sky. Optical observations \cite{Mig02} ruled out the 
possibility that the two objects are physically associated as it was assumed for a long time \citep{Cla75}. 

\section{Observations}

We have observed the X-ray binary Circinus X-1 during a time interval of almost 10 years, between 1996 and 2006, simultaneously at 4.8 and 8.6 GHz with 
the Australia Telescope Compact Array (ATCA). As primary calibrators we used PKS J0825-5010 (PKS B0823-500) for epochs 15 and 37 (see Table 1) and 
PKS J1939-6342 (PKS B1934-638) for the rest. PMN J1524-5903 (B1520-58) was used as secondary calibrator for the entire data set. Its positional 
uncertainty is, according to the ATCA calibrator catalog, between 100 and 250 mas. Throughout the paper we assumed conservatively the above upper limit as the 
systematic error in our determinations of positions. Standard calibration techniques were applied using the software \textsc{MIRIAD} \citep*{Sau95}.

The observations were carried out at different spatial resolutions, although very similar (see the array configuration column in Table 1 and the ATCA 
website\footnote{http://www.narrabri.atnf.csiro.au/observing/configs.html}). To aid an 
homogeneous analysis, in obtaining the final radio images we have used the same restoring beams (size and orientation) for the entire data set at each of 
the two observed frequencies, for each weighting scheme, natural and uniform. The size of these beams is in general less than 5 percent different than the 
sizes of the default synthesized beams. 
The exceptions are the observations from epoch 14 when the difference was up to 40 percents (array configuration 1.5D) and most of the observations 
from 1996 July for which the default synthesized beams were very elongated due to the short observational runs. 
Because of this limitation the radio images from the 1996 July epochs 3-7, 9 and 10 are very likely to contain artifacts. However, we note that the calibration 
process was successful and the {\it uv} data 
should not be significantly affected. The only special case is epoch 10, for which the phase-referencing calibration 
was only partially successful, but we are including it in the analysis for completeness.

The flux density within a run varied sometimes by up to 30 percent. In general the variations were smooth over the duration of the observing sessions and tests 
made by selecting only parts of the data showed 
that the maps are not affected significantly by this behaviour. However, when a radio flare happened during an observational run the flux density changed by 
almost a factor 2 on timescales of hours. In these few cases artifacts could be present in the radio maps (see section 3, paragraph 4).

As mentioned in the previous section, Circinus X-1 periodically flares at radio wavelengths. We used the ephemeris derived from the onset times of 21 well 
observed such outbursts spanning the past 28 years \citep{Nic07} to illustrate in Fig. 2 the orbital phase coverage of our observations. During the 10 
years of data we have swept the entire orbit of the system, even though more often than not only once for a given orbital phase, and gaps in the coverage are 
present.

\begin{table*}
 \centering
  \caption{Observational log. The table contains the ordinal number of the epoch of observations, the date of the observations, the configuration code of the 
ATCA antennae, the 
modified Julian Day of the beginning of the observations, the total time of the observing session, the orbital phase interval swept during the observational 
run (radio ephemeris from \citealt{Nic07}), 
the total flux densities (core+jet) at 4.8 and 8.6 GHz measured in the image plane. On epoch 39, at 8.6 GHz a 3 $\sigma$ upper limit is listed. On epoch 41 the 
target was detected at confidence levels of less than 3 $\sigma$.}
  \begin{tabular}{@{}cccccccc}
  \hline \hline
   Epoch  & Date & Array config. & MJD beginning & Total time [h] & Orbital phase & $F_{4.8}$ [mJy] & $F_{8.6}$ [mJy] \\
\hline
1 & 1996 July 01 & 6C & 50265.282 & 11.5 & 0.102 - 0.131 & 30.8 $\pm$ 7.0 & 24.6 $\pm$ 5.0 \\ 
2 & 1996 July 02 & 6C & 50266.291 & 8.4 & 0.163 - 0.184 & 31.5 $\pm$ 6.4 & 26.6 $\pm$ 5.1 \\
3 & 1996 July 03 & 6C & 50267.275 & 0.8 & 0.223 - 0.225 & 23.2 $\pm$ 5.3 & 24.1 $\pm$ 5.7 \\
4 & 1996 July 04 & 6C & 50268.276 & 0.8 & 0.283 - 0.285 & 23.7 $\pm$ 5.6 & 22.7 $\pm$ 5.0 \\
5 & 1996 July 05 & 6C & 50269.252 & 1.2 & 0.342 - 0.345 & 14.9 $\pm$ 3.3 & 10.6 $\pm$ 2.5 \\
6 & 1996 July 07 & 6C & 50271.247 & 1.2 & 0.463 - 0.466 & 5.9 $\pm$ 1.5 & 3.9 $\pm$ 0.8 \\
7 & 1996 July 08 & 6C & 50272.249 & 0.8 & 0.523 - 0.525 & 6.8 $\pm$ 1.8 & 5.9 $\pm$ 1.2 \\
8 & 1996 July 09 & 6C & 50273.187 & 4.4 & 0.580 - 0.591 & 9.4 $\pm$ 1.9 & 8.5 $\pm$ 1.3 \\
9 & 1996 July 10 & 6C & 50274.267 & 1.2 & 0.645 - 0.648 & 21.9 $\pm$ 4.9 & 19.4 $\pm$ 4.0 \\
10 & 1996 July 13 & 6C & 50277.172 & 0.8 & 0.821 - 0.823 & 8.2 $\pm$ 2.0 & 4.7 $\pm$ 1.1 \\
11 & 1998 February 05/06 & 6A & 50849.573 & 12.3 & 0.424 - 0.455 & 9.2 $\pm$ 1.6 & 6.8 $\pm$ 0.8 \\
12 & 1998 February 23/24 & 6B & 50867.555 & 12.3 & 0.512 - 0.543 & 10.8 $\pm$ 1.7 & 7.4 $\pm$ 0.9 \\
13 & 1998 October 03/04 & 6A & 51089.883 & 13.1 & 0.954 - 0.987 & 7.5 $\pm$ 1.0 & 4.0 $\pm$ 0.4 \\
14 & 1998 October 16/17 & 1.5D & 51102.829 & 9.2 & 0.736 - 0.759 & 6.0 $\pm$ 0.9 & 3.9 $\pm$ 0.3 \\
15 & 1998 October 29/30 & 6D & 51115.810 & 7.6 & 0.521 - 0.540 & 7.9 $\pm$ 1.3 & 4.1 $\pm$ 0.6 \\
16 & 1998 November 02 & 6D & 51119.100 & 7.6 & 0.720 - 0.739 & 5.0 $\pm$ 0.7 & 2.8 $\pm$ 0.2 \\
17 & 2000 February 05/06 & 6A & 51579.647 & 12.7 & 0.568 - 0.600 & 16.2 $\pm$ 2.2 & 8.9 $\pm$ 0.9 \\
18 & 2000 October  01 & 6A & 51818.081 & 8.8 & 0.987 - 0.009 & 21.1 $\pm$ 1.9 & 9.2 $\pm$ 0.6 \\
19 & 2000 October  07/08 & 6A & 51824.875 & 4.0 & 0.398 - 0.408 & 18.1 $\pm$ 2.2 & 8.3 $\pm$ 0.7 \\
20 & 2000 October  09/10 & 6A & 51826.950 & 11.5 & 0.524 - 0.553 & 20.0 $\pm$ 2.5 & 10.4 $\pm$ 0.8 \\
21 & 2000 October  14/15 & 6A & 51831.998 & 10.0 & 0.829 - 0.854 & 27.4 $\pm$ 3.3 & 14.5 $\pm$ 1.2 \\
22 & 2000 October  19 & 6C & 51836.067 & 8.8 & 0.075 - 0.097 & 34.1 $\pm$ 3.9 & 21.0 $\pm$ 1.5 \\
23 & 2000 October  20/21 & 6C & 51837.922 & 11.5 & 0.187 - 0.216 & 30.9 $\pm$ 4.2 & 18.7 $\pm$ 1.9 \\
24 & 2000 October  23 & 6C & 51840.185 & 6.4 & 0.324 - 0.340 & 25.1 $\pm$ 2.8 & 14.2 $\pm$ 1.0 \\
25 & 2000 October  25/26 & 6C & 51842.909 & 11.5 & 0.489 - 0.518 & 24.4 $\pm$ 3.0 & 13.9 $\pm$ 1.0 \\
26 & 2001 May 23 & 6F & 52052.277 & 10.7 & 0.151 - 0.178 & 19.2 $\pm$ 2.5 & 12.6 $\pm$ 1.1 \\
27 & 2001 May 25 & 6F & 52054.230 & 9.2 & 0.269 - 0.292 & 14.6 $\pm$ 1.9 & 9.5 $\pm$ 0.7 \\
28 & 2001 May 27 & 6F & 52056.235 & 9.2 & 0.390 - 0.413 & 16.9 $\pm$ 1.9 & 10.0 $\pm$ 0.7 \\
29 & 2001 May 29 & 6F & 52058.219 & 10.0 & 0.510 - 0.535 & 15.7 $\pm$ 2.0 & 9.6 $\pm$ 0.7 \\
30 & 2001 September 08 & 6B & 52160.190 & 6.4 & 0.677 - 0.693 & 9.3 $\pm$ 0.8 & 4.0 $\pm$ 0.3 \\
31 & 2002 December 02/03 & 6A & 52610.795 & 11.5 & 0.933 - 0.962 & 3.3 $\pm$ 0.2 & 0.7 $\pm$ 0.1 \\
32 & 2002 December 03/04 & 6A & 52611.813 & 11.9 & 0.994 - 0.024 & 3.7 $\pm$ 0.2 & 1.7 $\pm$ 0.1 \\
33 & 2002 December 04/05 & 6A & 52612.848 & 11.5 & 0.057 - 0.086 & 14.9 $\pm$ 1.4 & 15.2 $\pm$ 1.8 \\
34 & 2002 December 05/06 & 6A & 52613.850 & 11.9 & 0.117 - 0.147 & 11.7 $\pm$ 1.9 & 6.8 $\pm$ 1.1 \\
35 & 2002 December 06/07 & 6A & 52614.813 & 12.3 & 0.176 - 0.207 & 6.9 $\pm$ 0.9 & 3.7 $\pm$ 0.5 \\
36 & 2002 December 07/08 & 6A & 52615.811 & 11.5 & 0.236 - 0.265 & 4.3 $\pm$ 0.4 & 1.8 $\pm$ 0.1 \\
37 & 2002 December 08/09 & 6A & 52616.822 & 10.0 & 0.297 - 0.322 & 2.8 $\pm$ 0.3 & 0.9 $\pm$ 0.1 \\
38 & 2003 December 23/24 & 6A & 52996.829 & 11.5 & 0.285 - 0.313 & 1.7 $\pm$ 0.2 & 0.4 $\pm$ 0.1 \\
39 & 2005 April 06 & 6A & 53466.476 & 11.5 & 0.699 - 0.727 & 1.4 $\pm$ 0.1 & $<$ 0.12 (3 $\sigma$)  \\
40 & 2005 June 17 & 6B & 53538.266 & 11.5 & 0.043 - 0.071 & 42.8 $\pm$ 8.0 & 40.0 $\pm$ 7.7 \\
41 & 2006 March 22/23 & 6C & 53816.562 & 11.5 & 0.882 - 0.910 & 0.09 (1.4 $\sigma$) & 0.08 (1.6 $\sigma$) \\
\hline \hline
\end{tabular}
\end{table*}

\section{``Secular'' evolution}

Fig. 1 shows the ``secular'' light curves of Circinus X-1 in radio and X-ray, between 1996-2006. The averaged X-ray output in the 2-10 keV band was constant 
around 75 counts s$^{-1}$ ($\sim$ 1 Crab) up to the beginning of 2000 when it started to decrease relatively fast reaching only a few tens of counts 
s$^{-1}$ by the end of 2003. The radio fluxes plotted (presented in columns 6 and 7 of Table 1) are determined in the image-plane 
and are total flux densities, in the sense that contain both the contribution of the core and other radio emitting regions (jet features, knots) associated 
with the system. The radio data offer a totally different picture than in the 1970-1980s when the flares usually reached flux density levels of up to 1 Jy. 
From 1996 till 2006 the outbursts only peaked at a few tens of mJy. Admittedly, we only observed a few times close to the orbital phase 0.0, but the flux 
densities measured at other phases further support the evidence that indeed dramatic changes took place in the system since the 1980s. The true 
nature of these changes is unclear.  

At a shorter temporal scale, whether the decrease in the X-ray output observed starting with 2000 can be correlated with a change in the radio behaviour 
is hard to assess with confidence. On the one hand it does seem that the radio flux density levels became lower by comparison with those observed at similar orbital 
phases before, however the limited number of observations cannot exclude the possibility of this being an observational bias. On the other hand the flare observed 
on 2005 June 17 (epoch 40, Table 1) had the highest flux densities in the whole data set, although this might just as well be associated with an increase in the 
radio activity of Circinus X-1 as was observed in the last few years \citep{Fen05,Nic07,Del07}.

\begin{figure*}
  \includegraphics[scale=0.30]{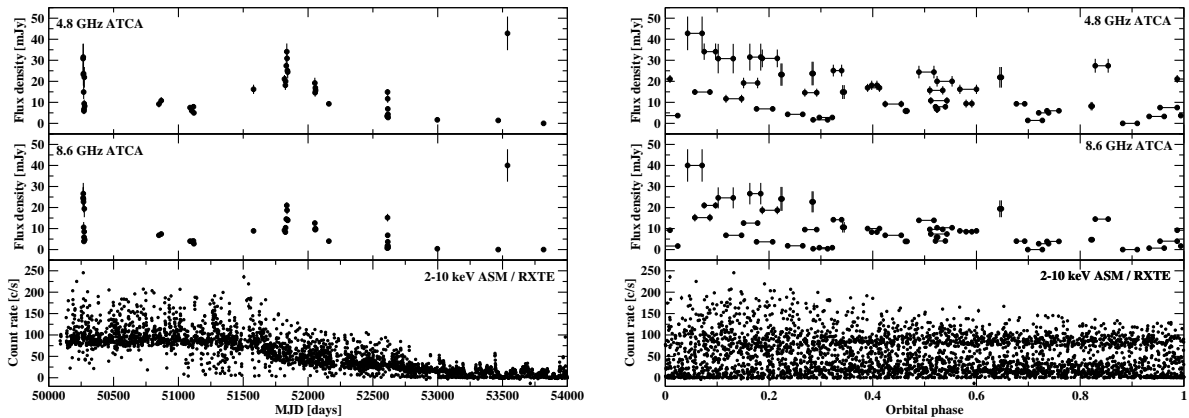}
  \caption{``Secular'' evolution of Circinus X-1. {\it Left, Top and Middle}: Variation of the total flux density between 1996-2006 at 4.8 and 8.6 
GHz from the ATCA radio data. {\it Left, Bottom}: 2-10 keV X-ray light curve over the same time interval from ASM/RXTE. {\it Right, 
Top and Middle}: Variation of the total flux density with respect to the orbital phase of the system at 4.8 and 8.6 GHz from the ATCA radio data. 
{\it Right, Bottom}: Changes in the 2-10 keV X-ray emission as a function of the orbital phase as detected by ASM/RXTE. Radio ephemeris from \citealt{Nic07}.}
\end{figure*}

\begin{figure}
  \includegraphics[angle=-90,scale=0.33]{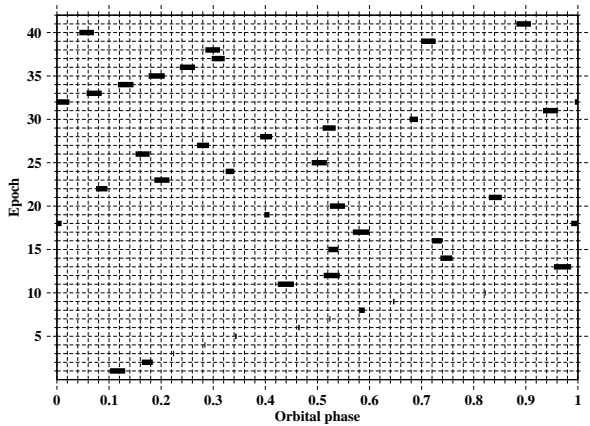}
  \caption{Orbital phase coverage of each epoch of observations (see Table 1). Radio ephemeris from \citealt{Nic07}.}
\end{figure}

Fig. 3 and Fig. 4 present the radio images of Circinus X-1 at 4.8 and 8.6 GHz. In most of the images a jet-like structure is evident towards SE. This is 
interpreted as the approaching jet in a system (a microquasar; e.g. \citealt{Mir99}) in which the ejection of matter is along a direction close to the line of 
sight (e.g. \citealt{Fen98,Fen04}). 
This arcsec scale jet has roughly the same position angle as the observed arcmin scale jet (e.g. \citealt{Ste93}) and might constitute part 
of the channel through which the energy is transferred from the core to the radio nebula \citep{Tud06}. In the same frame-work 
of the microquasar interpretation, at a few epochs (e.g. 23, 25, 28) the receding jet was identified as the excess of emission towards NW. 

Some of the observations close to phase 0.0 (images A,B - 18, 22, 32, 33 , 40 in Fig. 3 and Fig. 4) show a significant different orientation of the jet axis 
than the ``normal'' NW-SE direction. This is very likely an artifact, resulting from variability of the source during the aperture synthesis. Test maps 
made in these cases with {\it uv}-plane selected data outside the instants of the outbursts, show that the 
radio emission has a NW-SE orientation. In order to accommodate the radio observations, which require a small angle between the jet axis and the line of 
sight \citep{Fen04}, with the optical and X-ray data, which prefer a larger angle between the two \citep{Jon07,Sch08,Iar08}, it was suggested \citep{Iar08} 
that the jet is precessing. Although the data analyzed here cannot entirely rule out this possibility, we did not find any unequivocal evidence for 
significant changes in the orientation of the jet axis with respect to the line of sight.

The radio images between 1996-2006 reveal a system in which the variations in the morphology and brightness of the radio features are characterized by 
timescales of days. Individual, well defined emitting regions are seen sometimes as far as 10 arcsec from the core (0.05 $r$ pc, with $r$ the distance to 
Circinus X-1 expressed in kpc), along a NW-SE direction (e.g. images A - 35, 39). Unfortunately, 
the limited amount of data (mainly the lack of enough closely spaced observations) and the magnitude of the systematic errors didn't allow a confident 
measurement of the proper motion of the various features. 

In addition to the usual radio flares close to the orbital phase 0.0, we have identified outbursts also near the orbital phase 0.5 at epochs 15 and 20 (Fig. 5). 
Strong evidence for similar outbursts was also present at epochs 8, 11 and 25. This has been noted just once previously in the literature \citep{Fen97}. 
As is often the case for the flares close to the orbital phase 0.0, these outbursts also seem to appear preferentially as a series 
of flares, with a time separation of a few hours. Their relative amplitudes look similar, independent of their association with the periastron or apastron 
passage (Figs. 5, 6). We stress that according to the ephemeris used \citep{Nic07}, orbital phase 0.0 (which is interpreted as periastron passage) 
corresponds to the 
moment of the onset, not peak, of the flare. Although given the orbital coverage we cannot totally rule out observational biases (i.e. flares at other orbital 
phases), from our sample we can 
conclude that all the flaring events were associated with the orbital 
phases 0.0 and 0.5 (within a region of $\pm$ 0.1 in phase) and that this correlation suggests the existence of a causal relation between the increase in flux 
density and these particular positions along the orbit. 

As remarked before, the outbursts near orbital phase 0.0 are interpreted as episodes of enhanced accretion at the periastron 
passage \citep{Mur80}. The accreting material is provided via Roche lobe overflow and wind accretion from the massive star (if indeed the companion is a 
supergiant as suggested by \citealt{Mur80} and more recently by \citealt{Jon07}). The explanation for the radio flares close to the orbital phase 0.5 
is less clear. We suggest that the wind accretion is responsible for producing them. The wind accretion rate is given by (e.g. \citealt{Bon44}):

\begin{equation}
\dot{M} \propto G^{2} M^{2} \rho_{w} v_{rel}^{-3}
\end{equation}
where $G$ is the gravitational constant, $M$ is the mass of the compact object, $\rho_{w}$ is the density of the stellar wind and $v_{rel}$ is the relative 
velocity between the compact object and the stellar wind. In an eccentric binary system, assuming homogeneous wind, the accretion rate has two peaks: one near 
the periastron, the other 
where the decrease in the relative velocity overcompensates for the decrease in the density of the wind. The exact position and width of this second 
maximum depend on the properties of the system, notably on the characteristics of the stellar wind (e.g. \citealt{Mar95}). In this frame-work the multiple 
flaring components observed during an outburst can naturally be explained by invoking inhomogeneities present in the wind. However, given the uncertainties 
in the parameters characterizing the system, a more quantitative test of the hypothesis is not viable at the present time. Alternatively, the flares close 
to the orbital phase 0.5 might be related to the disc settling down into another increased accretion rate state after a period of reduced or suppressed 
accretion or settling down into a radio-loud state after a radio-quiet one (cf. black holes; e.g. \citealt{Fen06}).  

\begin{figure*}
  \includegraphics[scale=0.2]{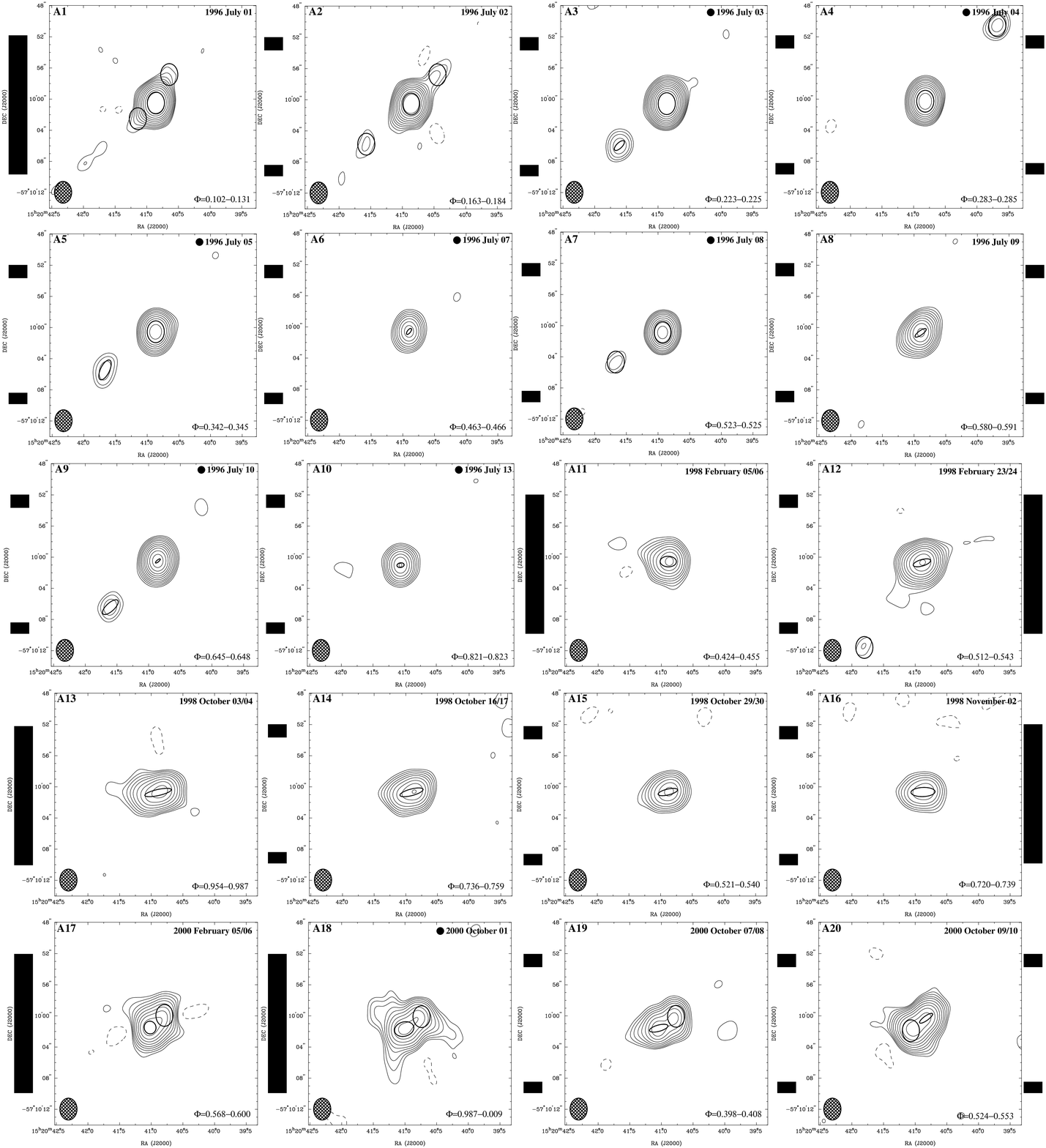}
  \caption{Radio images (natural weightings) of Circinus X-1 at 4.8 GHz. The contour lines are at -2.8, 2.8, 4, 5.6, 8, 11, 16, 23, 32, 45, 64, 90 $\times$ 
the rms noise at 
each epoch. The size of the restoring beam is 2.8 $\times$ 2.2 arcsec$^2$ at PA=0\fdg0. Superimposed are the results of the image-plane fitting process 
(see Appendix A). A filled circle next to the date of observation warns that the map might contain artifacts (see section 2). The fitted unresolved Gaussian 
components are represented as ellipses with the size and orientation of 
the restoring beam. The ordinal number of the 
epoch of observation and the corresponding date and orbital phase are shown on each map. Observations made at 
relatively short time intervals (i.e. in a single campaign, often within a single binary orbit) are organized in subsets separated by black vertical lines. 
Epoch 41 is not shown.}
\end{figure*}

\begin{figure*}
  \includegraphics[scale=0.2]{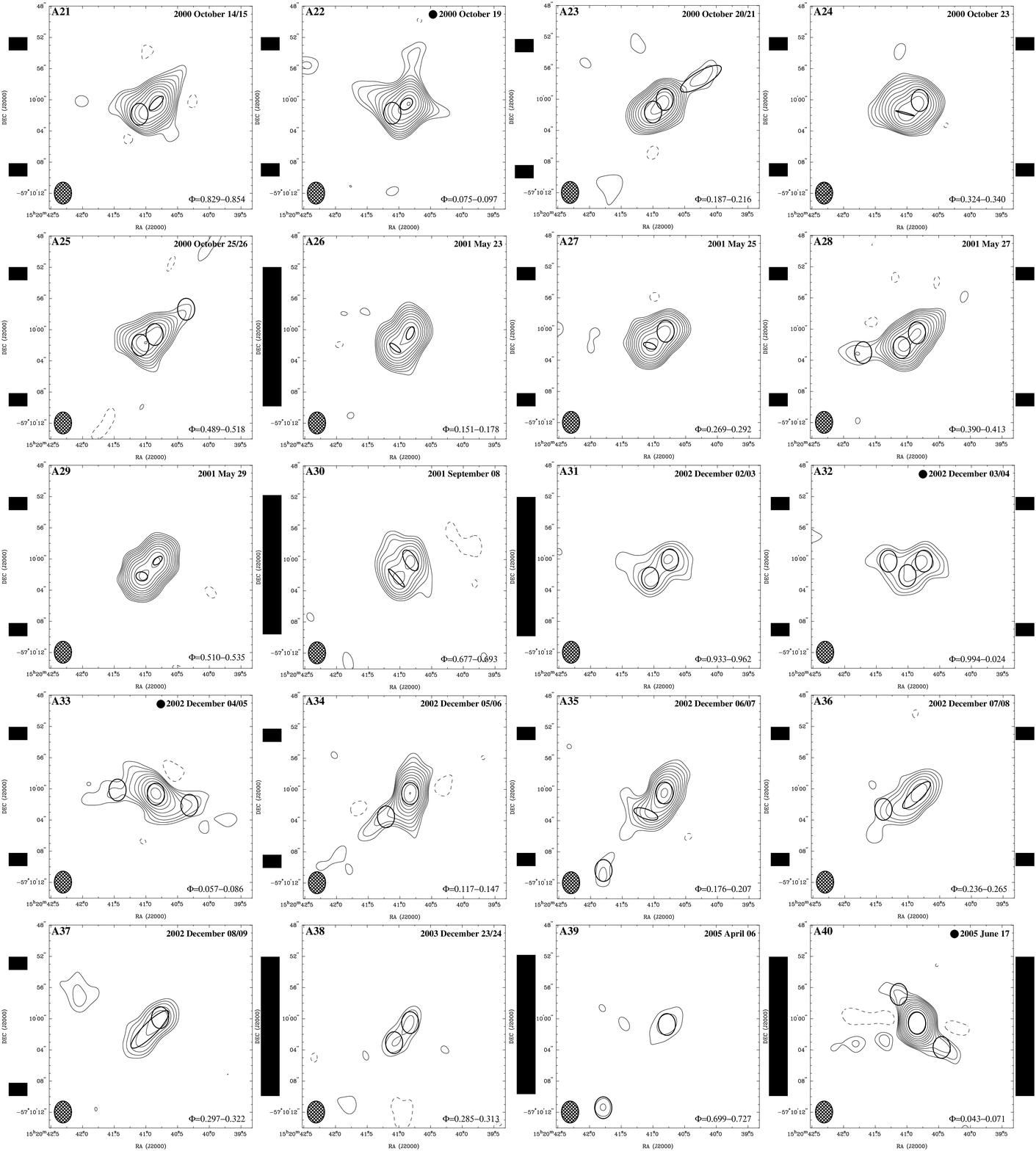}
  \contcaption{}
\end{figure*}

\begin{figure*}
  \includegraphics[scale=0.2]{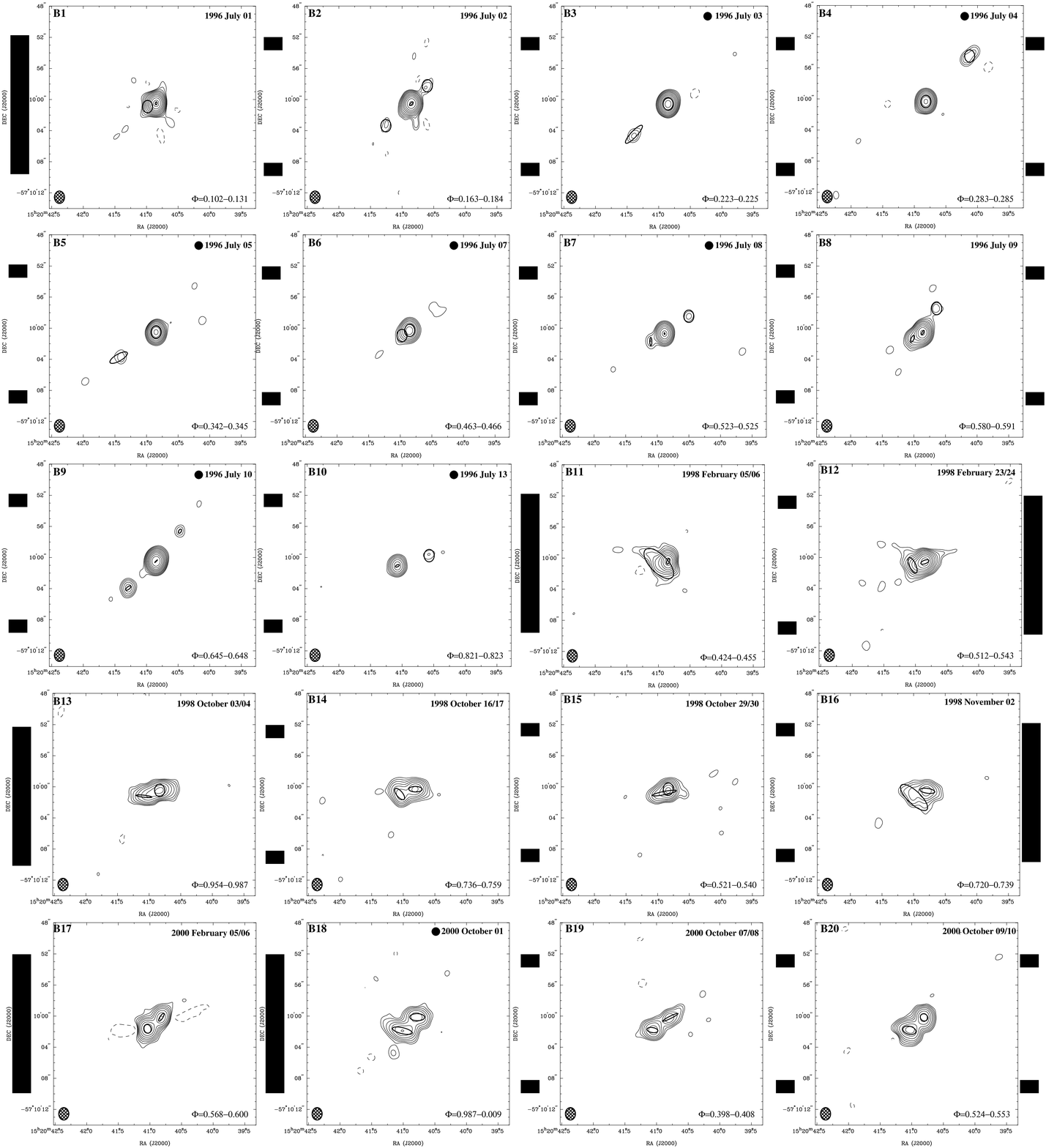}
  \caption{Radio images (natural weightings) of Circinus X-1 at 8.6 GHz. The contour lines are at -2.8, 2.8, 4, 5.6, 8, 11, 16, 23, 32, 45, 64, 90 $\times$ 
the rms noise at 
each epoch. The size of the restoring beam is 1.6 $\times$ 1.3 arcsec$^2$ at PA=0\fdg0. Superimposed are the results of the image-plane fitting process 
(see Appendix A). A filled circle next to the date of observation warns that the map might contain artifacts (see section 2). The fitted unresolved Gaussian 
components are represented as ellipses with the size and orientation of 
the restoring beam. The ordinal number of the 
epoch of observation and the corresponding date and orbital phase are shown on each map. Observations made at 
relatively short time intervals (i.e. in a single campaign, often within a single binary orbit) are organized in subsets separated by black vertical lines. 
Epoch 41 is not shown.}
\end{figure*}

\begin{figure*}
  \includegraphics[scale=0.2]{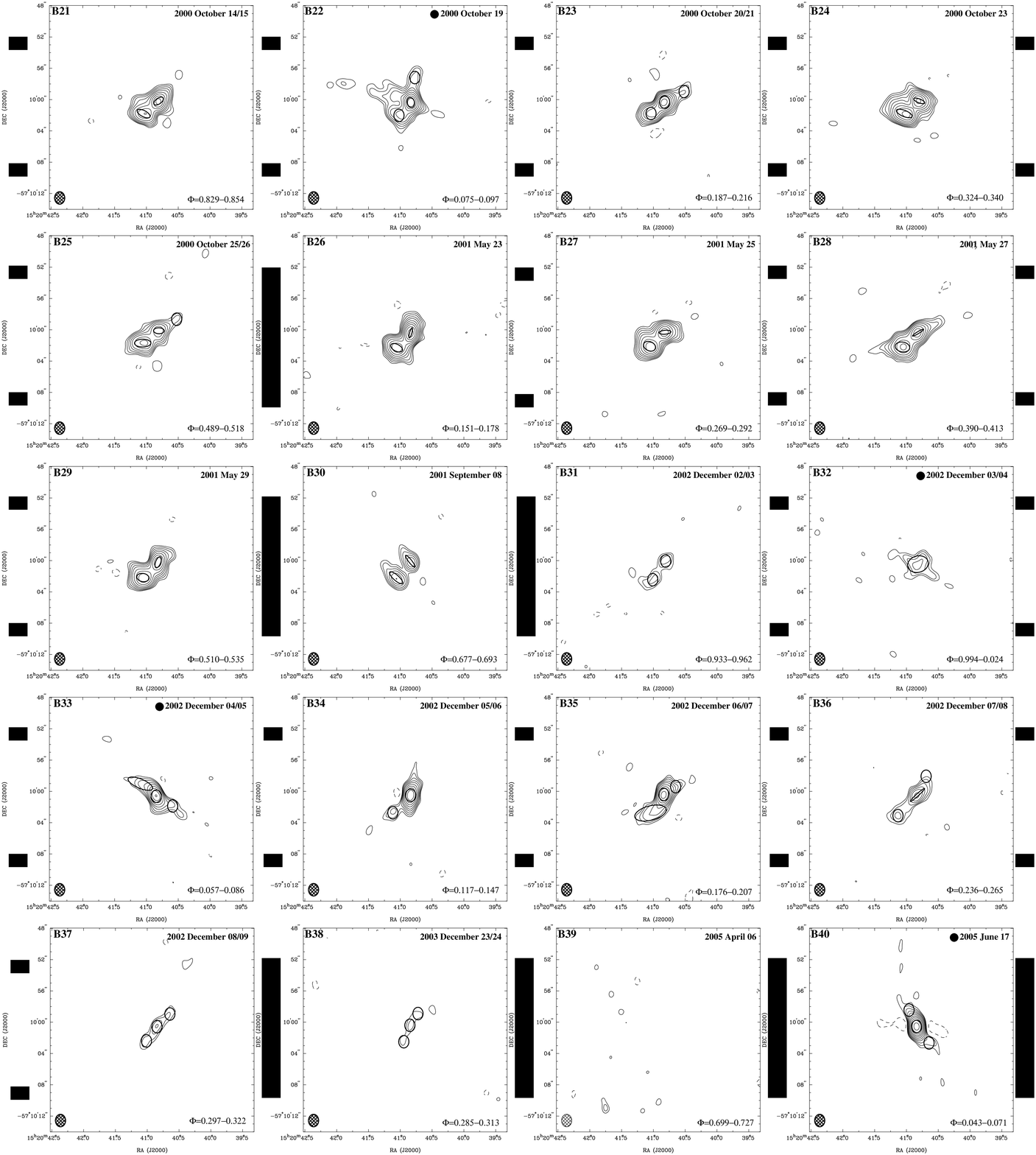}
  \contcaption{}
\end{figure*}

\begin{figure*}
  \includegraphics[scale=0.3]{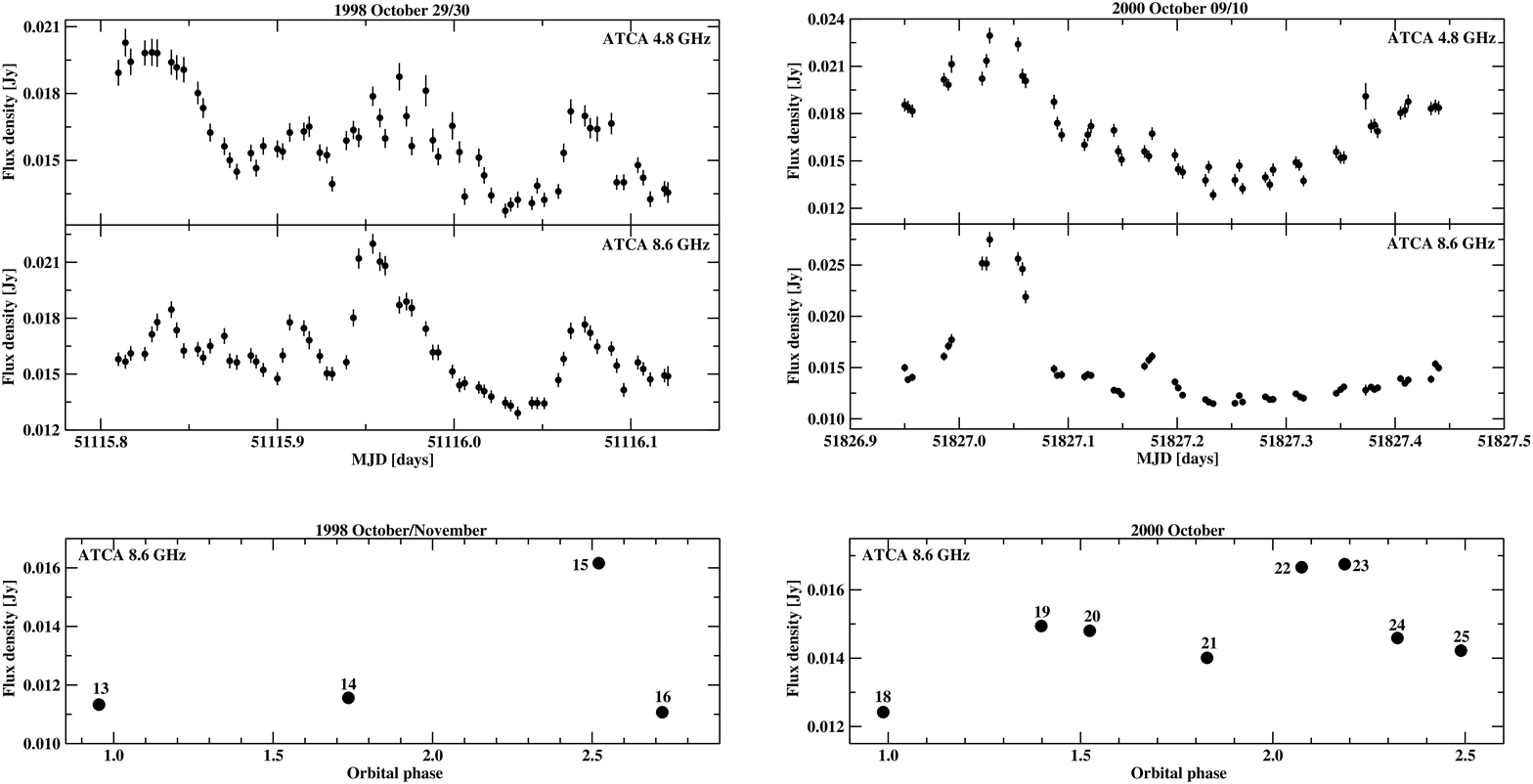}
  \caption{{\it Top:} Radio flares near orbital phase 0.5. The plots show 4.8 and 8.6 GHz ATCA 5 min averaged {\it uv}-plane light curves of epoch 15 
({\it Left}) and epoch 20 ({\it Right}). Multiple flaring events are visible, more clearly at the higher frequency. {\it Bottom:} 8.6 GHz ATCA 1 day 
averaged {\it uv}-plane light curves showing the longer timescales flux variations, near the times of flaring activity. The numbers next to the points 
correspond to the ordinal number of the epoch of observations. The formal errors in the flux measurements are smaller than the size of the points.}
\end{figure*}

\begin{figure*}
  \includegraphics[scale=0.3]{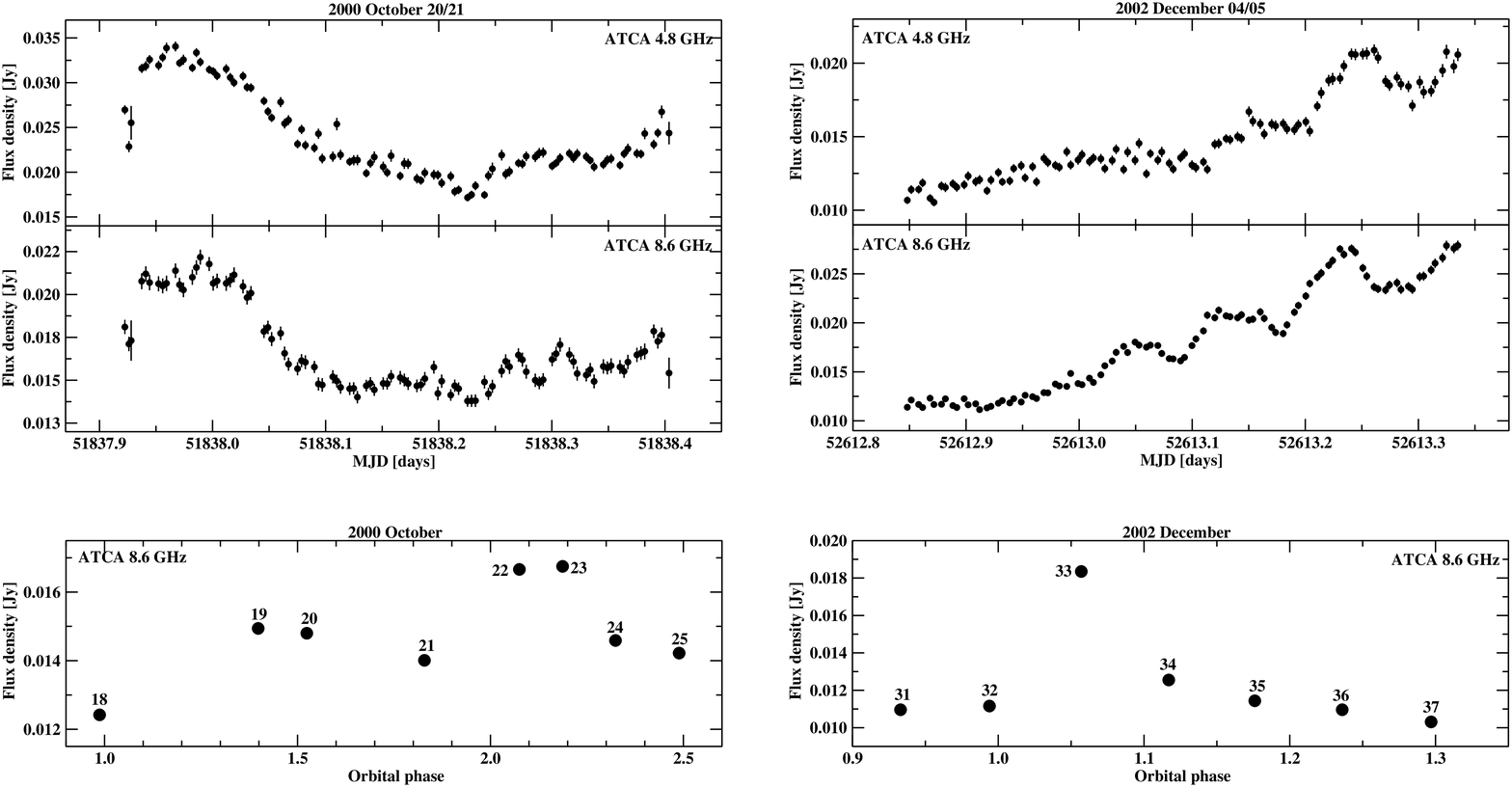}
  \caption{{\it Top:} Radio flares near orbital phase 0.0. The plots show 4.8 and 8.6 GHz ATCA 5 min averaged {\it uv}-plane light curves of epoch 23 
({\it Left}) and epoch 33 ({\it Right}). Multiple flaring events are visible, more clearly at the higher frequency. {\it Bottom:} 8.6 GHz ATCA 1 day 
averaged {\it uv}-plane light curves showing the longer timescales flux variations, near the times of flaring activity.  The numbers next to the points 
correspond to the ordinal number of the epoch of observations. The formal errors in the flux measurements are smaller than the size of the points.}
\end{figure*}

In order to provide a more quantitative record of the evolution of Circinus X-1 over the period 1996-2006 we have fitted the data in the image-plane using 
the task \textsc{IMFIT} in \textsc{MIRIAD}. The number of components (admittedly subjective) used in the 
fitting process varied between one and three, according to the appearance of the radio images. As a general rule, a new component was added to the fitting 
only in the case the residual map (obtained by subtracting the model map from the original, CLEAN-ed map) was showing compact emission at more than 3 
$\sigma$ level. All the components were refitted then simultaneously. The results are presented in Appendix A and Figures 3 and 4. To check the reliability 
of the image-plane fittings, we have also performed 
{\it uv}-plane fittings of the data at 8.6 GHz using the software \textsc{DIFMAP}. The two different approaches gave similar results, within the errors. 

\section{Ultra-relativistic outflow}

Using a sub-sample of our data at 4.8 GHz, \cite{Fen04} found evidence for an underlying relativistic flow in the arcsec scale jet of Circinus X-1. 
We have reanalyzed their 
data set in this context, taking also into account the simultaneous 8.6 GHz data. Only the observations made in 2000 October, 2001 May and 2002 December 
were suitable for such a study due to their dense temporal coverage (see Table 1). These are the same data presented in \cite{Fen04}. The 1996 July subset was 
disqualified because of the many short individual 
observations and the likely artifacts that might be thus generated in the images. For each of the three subsets mentioned the data were selected such as to 
ensure a quasi-identical {\it uv}-plane coverage for the observations. For 2000 October and 2001 May we used for the image-plane fitting uniformly 
weighted radio maps since they revealed with a higher spatial resolution the structure of the radio emission. For 2002 December this was not possible 
because on many epochs Circinus X-1 was too weak and so we used the naturally weighted radio maps 
instead. The data from epoch 19 (2000 October 7/8) were not included in the analysis as a consequence of the limited common {\it uv}-plane coverage with the 
other observations from the same month. On a few epochs in 2002 December, the component 2 was not detected or it was impossible to identify with certainty. 
The approximate location of the components fitted to the data is indicated in Fig. 7 along with the 
corresponding light curves. Assuming that the peaks of the flux density of component 1 (the core) and component 2 observed at timescales of days are causally 
related (see for instance Fig. 8 showing the successive variations in the relative brightness of the different features of the radio emission) then the radio 
brightening of the core and of the knots further away appear to happen with a delay of days. The X-ray burst peaks in the 
same time interval and at least for some events evidence is mounting that the radio and X-ray outputs are correlated \citep{Sol07}. 
This phenomenology can be interpreted as evidence for a flow propagating outward from the core. If that is the case, then the outflow is indeed 
relativistic in Circinus X-1. The apparent velocity of the outflow in the plane of the sky can be expressed as:
\begin{equation}
\beta_{app}=\frac{v_{app}}{c} \simeq 5.8 \left(\frac{\phi}{arcsec}\right) \left(\frac{\Delta t}{day}\right)^{-1} \left(\frac{r}{kpc}\right)
\end{equation}
where $c$ is the speed of light, $\phi$ is the angular distance traveled by the flow in the time interval $\Delta t$ and $r$ is the distance to Circinus X-1. 
Allowing for different possible identifications of the succession of the peaks of the flux densities, estimations of the apparent velocity of the outflow 
can be made (Table 2). The formal error in column 5 of Table 2 is up to 30 percent. The most confident results are obtained for 2000 October, when relatively 
more observations are available and the identification of the peaks in the light curves is more evident. A minimum apparent velocity of around $3r$ seems to be 
favoured by this data. The other two data subsets tend to suggest a slightly higher value. Even though the apparent velocities of the 
outflow might have different values during different outburst events, it is likely that this interval of variation cannot be very large. In conclusion, given 
the limited amount of information that could be extracted from the data, we consider that there is enough circumstantial evidence that a minimum apparent 
velocity of 
around 3 $\pm$ 1 $r$ can be confidently assigned to the outflow in Circinus X-1. For a distance $r$ between 4 and 10.5 $kpc$ (e.g. \citealt{Jon04,Iar05}) 
this confirms the status of Circinus X-1 as harbouring the most relativistic outflow discovered so far in the Milky Way \citep{Fen04}, with $\beta_{app} \geq 12$. 

\begin{figure}
  \includegraphics[scale=0.32]{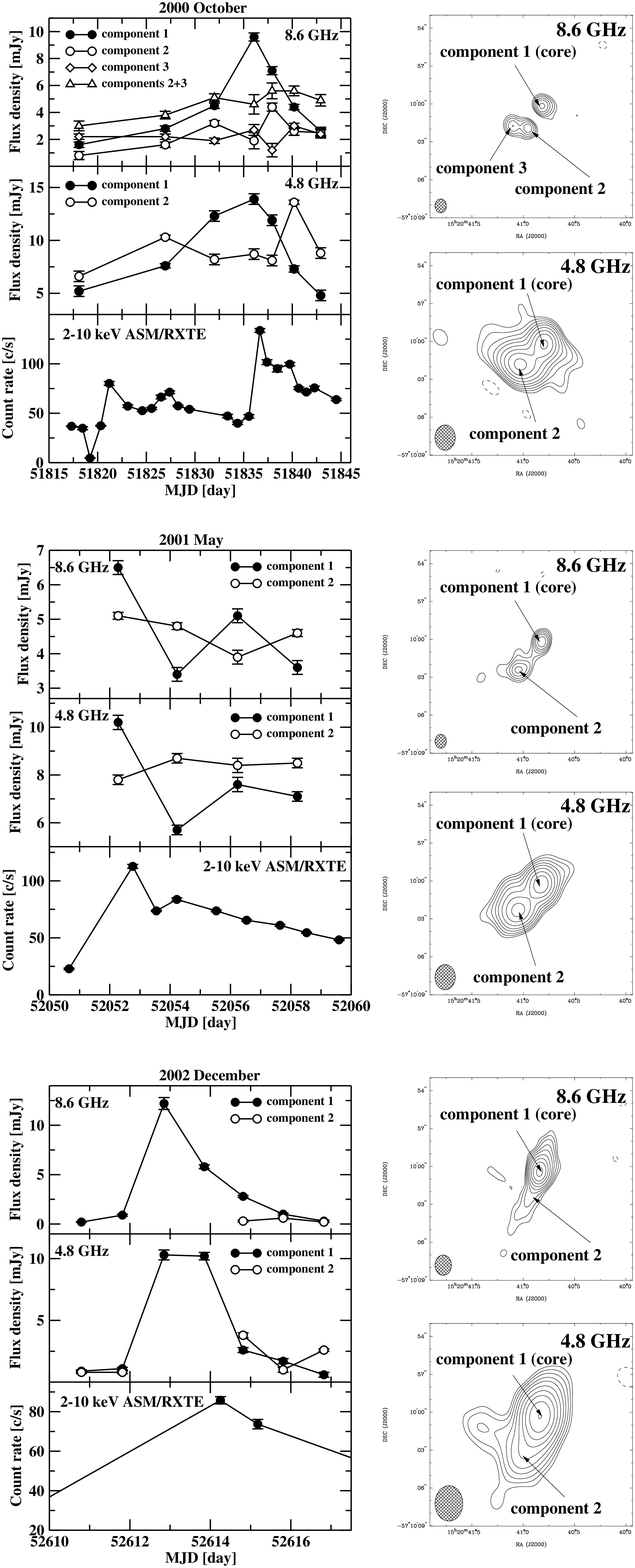}
  \caption{8.6 and 4.8 GHz ATCA radio light curves and 2-10 keV ASM/RXTE daily averaged X-ray light curves of Circinus X-1 for 2000 October, 
2001 May and 2002 December (a few epochs not included, see text). The fittings 
were done in the image plane, after selecting the same {\it uv}-plane coverage for all the observations in each of the three sessions. Uniformly weighted maps 
were used for 2000 October and 2001 May, and naturally weighted maps for 2002 December. The approximate location of the fitted components whose evolution was 
tracked is indicated. The successive brightening of different components happens within days.}
\end{figure}

\begin{figure}
  \includegraphics[scale=0.16]{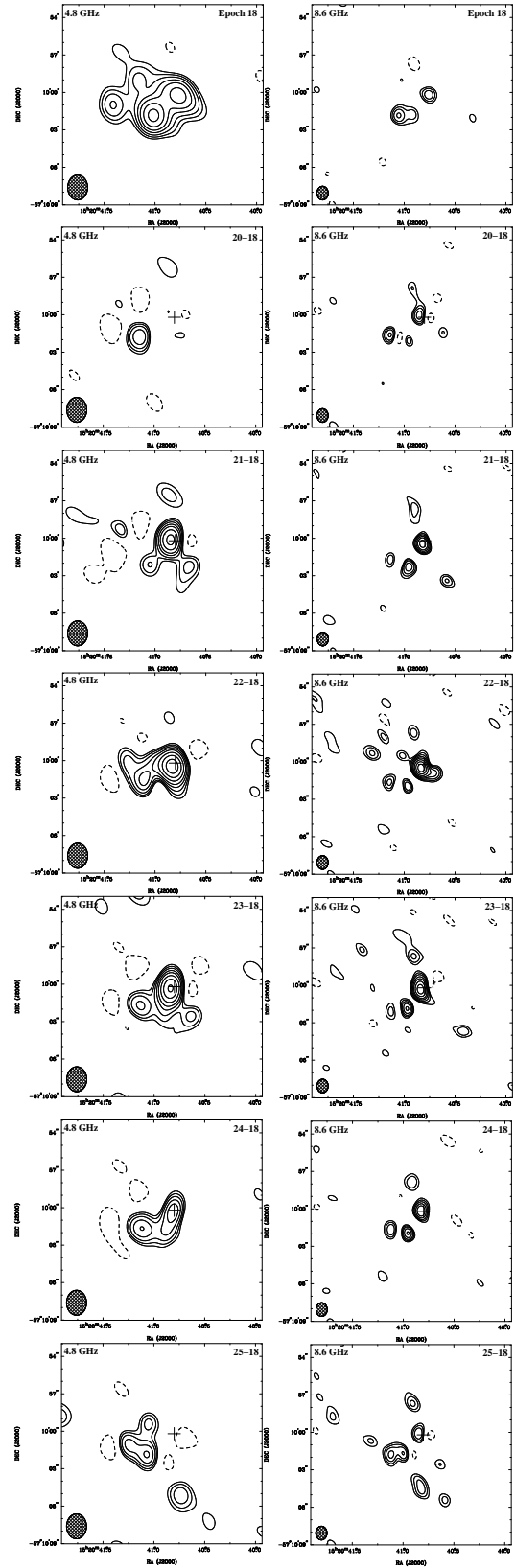}
  \caption{Difference radio maps for 2000 October obtained by subtracting the map on 2000 October 1 (epoch 18) from the maps corresponding to 
subsequent epochs. The subtraction was done in the image-plane using uniformly weighted images and is model independent. The plus sign shows the approximate 
position of the core.}
\end{figure}

\begin{table}
 \centering
  \caption{The apparent velocity $\beta_{app}$ of the outflow. Different possibilities for the identification of the time of the leading and following peaks 
are listed (see Fig. 7). $\phi$ is the angular distance between component 1 (core) and component 2. $\Delta t$ is the time delay between the leading and 
following peaks. $r$ is the distance to Circinus X-1 (between 4 and 10.5 $kpc$, e.g. \citealt{Jon04,Iar05}).}
   \begin{tabular}{@{}lcccc}
  \hline \hline
   Leading peak & Following peak & $\phi$ & $\Delta$t & $\beta_{app}/r$ \\

   [MJD] & [MJD] & [arcsec] & [day] & [kpc$^{-1}$] \\
\hline
2000 October & & & & \\
\hline
51836.067 & 51837.922 & 2.3 & 1.86 & 7.2 \\
51836.067 & 51840.185 & 2.3 & 4.12 & 3.2 \\
\hline
2001 May & & & & \\
\hline
52052.277 & 52054.230 & 2.85 & 1.95 & 8.5 \\
52052.277 & 52058.219 & 2.85 & 5.94 & 2.8 \\ 
52056.235 & 52058.219 & 2.85 & 1.98 & 8.4 \\
\hline
2002 December & & & & \\
\hline
52612.848 & 52614.813 & 3.6 & 1.96 & 10.6 \\
52612.848 & 52616.822 & 3.6 & 3.97 & 5.3 \\
52613.850 & 52614.813 & 3.6 & 0.96 & 21.7 \\
52613.850 & 52616.822 & 3.6 & 2.97 & 7.0 \\
\hline \hline
\end{tabular}
\end{table}

\section{System proper motion}

Circinus X-1 complex is on the plane of the sky just $\sim$ 10 arcmin away from the boundary of the supernova remnant SNR G321.9-0.3. This offered 
circumstantial evidence 
for the so called ``runaway binary'' scenario in which the binary system and the supernova remnant are physically associated \citep*{Cla75}. However, 
optical observations taken 8.6 years apart \citep{Mig02} placed a 3 $\sigma$ upper limit to the proper motion of the source of $\sim$ 5 mas yr$^{-1}$, 
much less than the expected value between 15 and 75 mas yr$^{-1}$ predicted by the ``runaway binary'' hypothesis.

In Fig. 9 (top) we plot the positions of the core of Circinus X-1 between 1996-2006 as determined from the 8.6 GHz data. The observations from epoch 10 
were excluded from the sample due to the phase-referencing problems encountered in calibrating this particular data set, which resulted in 
unaccountable errors in the position of the object, as can be actually clearly seen in the radio maps not only at 8.6, but also 4.8 GHz 
(Figs. 3 and 4, images A10 and B10). The diagram shows an apparent tendency of the points to align along a NW-SE direction (this tendency is also 
observed at 4.8 GHz). However, the position of the object ``jumps'' randomly from one epoch to another, with no 
preferential direction. To test if this could be explained by errors in the phase-referencing process, we used a compact source $\sim$ 7 $\arcmin$ 
away from Circinus X-1, designated J1520.6-571 by \cite{Fen98} which was observed during the runs starting with epoch 11. After phase-referencing it 
with respect to PMN J1524-5903 we fitted the data in the image plane. The resulting positions are reported in Fig. 9 (bottom). The lack in this case of 
a similar trend in the distribution of positions as observed for Circinus X-1 strongly suggests that the phase-referencing process is not responsible for 
the tendency (or, more conservatively, is not the dominant factor). Instead, given the magnitude of the errors and the fact that the bulk of the radio 
emission in Circinus X-1 is oriented on a NW-SE direction (the jet axis), it is very likely that this tendency is an artifact of the fitting process, 
due to different jet structures at different epochs. 
Therefore, our roughly estimated upper limit to the proper motion of Circinus X-1 of 50 mas yr$^{-1}$ is extremely conservative. This nevertheless places 
an independent constraint on the velocity of the system in the plane of the sky of $v_{system} <$ 240 $r$ km s$^{-1}$, where $r$ is the distance to 
Circinus X-1 expressed in kpc.

The errors in the position of the core are dominated at most of the epochs by the positional error of 
the phase-referencing calibrator (the exact value of which being unknown and assumed here to be the upper limit quoted in the catalog; see section 2). 
Future long-term observations of Circinus X-1 combined with a better knowledge of the systematic errors due to the calibrator will enable to further 
constrain the proper motion down to, and ultimately even better than the presently available optical limits.

\begin{figure}
  \includegraphics[scale=0.3]{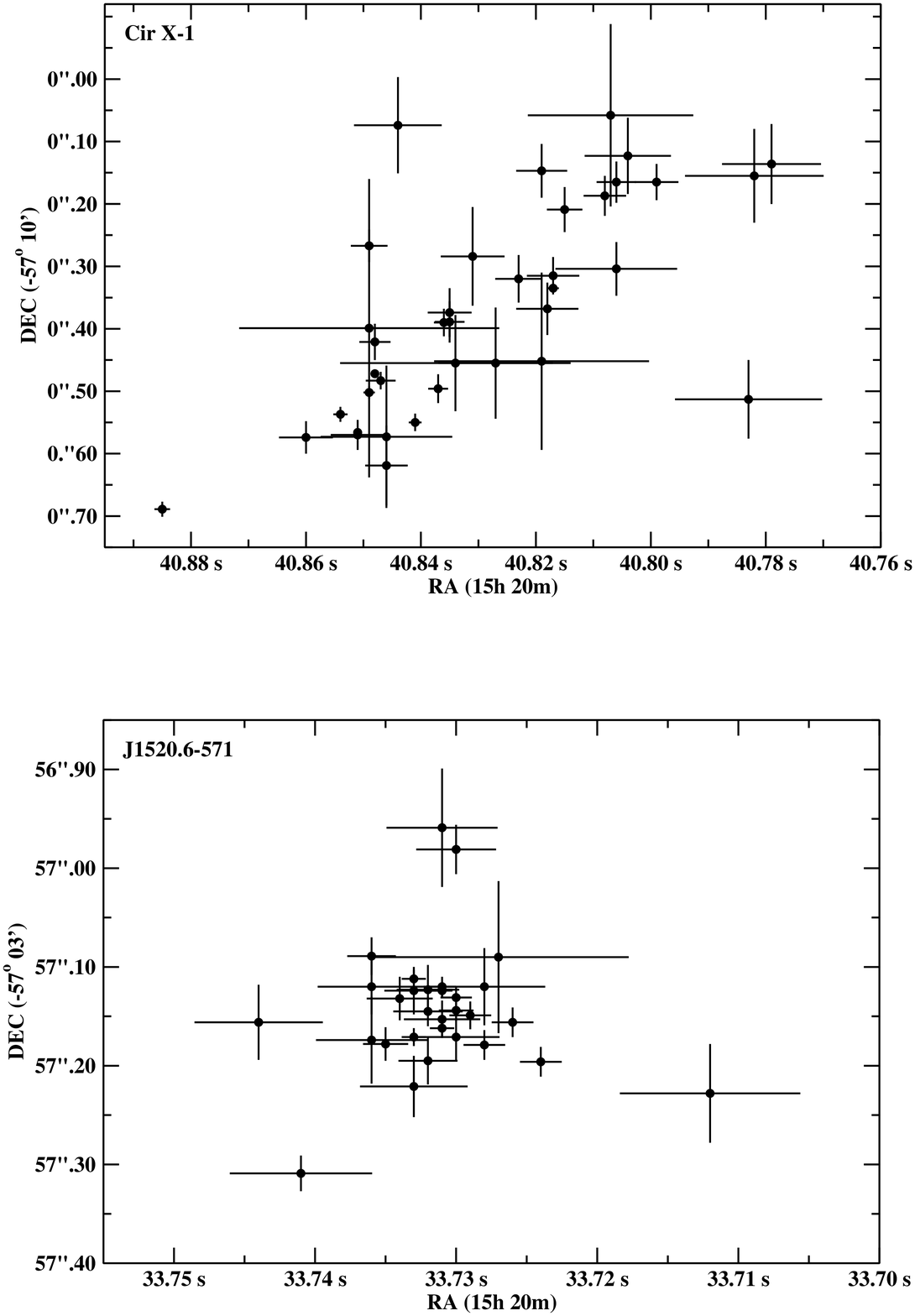}
  \caption{Positions of the core of Circinus X-1 ({\it top}) and J1520-571 ({\it bottom}) between 1996-2006. The coordinates were obtained via image 
plane fitting of the 8.6 GHz data. For an easier comparison the error bars do not contain the systematic errors associated with the position of the 
phase-referencing calibrator. Epoch 10 was excluded from the analysis (see text). The larger scatter in the fitted position of Circinus X-1 is along 
the jet axis and so is unlikely to be real.}
\end{figure}

\section{Spectrum}

The global spectral index of Circinus X-1, determined between 4.8 and 8.6 GHz, over the 10 years period of observations shows preferentially a steep 
spectrum with mean 
value and standard deviation (characterizing the scatter over the entire observed period) $\alpha=-0.9 \pm 0.6$ $(F_{\nu} \propto \nu^{\alpha})$ (Fig. 10). 
This is, unsurprisingly, indicative of optically thin synchrotron emission.

In the observational runs in which flares were detected, clear evidence was found, both close to orbital phases 0.0 and 0.5, for a significant flattening 
of the spectrum during these outbursts. This behaviour is due to the fact that the flux density peaks first at 8.6 GHz, then at 4.8 GHz. The time lag is 
of the order of a few hours (see Figs. 5 and 6). The peak flux density is higher at 8.6 GHz than at 4.8 GHz (however see epoch 23 in Fig. 6). 
Such phenomenology was pointed out before, notably by \cite{Hay78}. 

These properties of the flares are roughly consistent both with a model in which the synchrotron radiation is produced in a cloud of relativistic particles 
expanding adiabatically \citep{Laa66,Hay80}, and with the internal shock model, in which 
the particles are accelerated in successive shocks produced in a quasi-continuous jet \citep*{Kai00,Vad03,Fen04b}. The data do not clearly rule out either of 
the two. Free-free absorption in the wind of the companion might also be at work in the system thus complicating any attempt to an interpretation at the moment. 

\begin{figure}
  \includegraphics[angle=-90,scale=0.3]{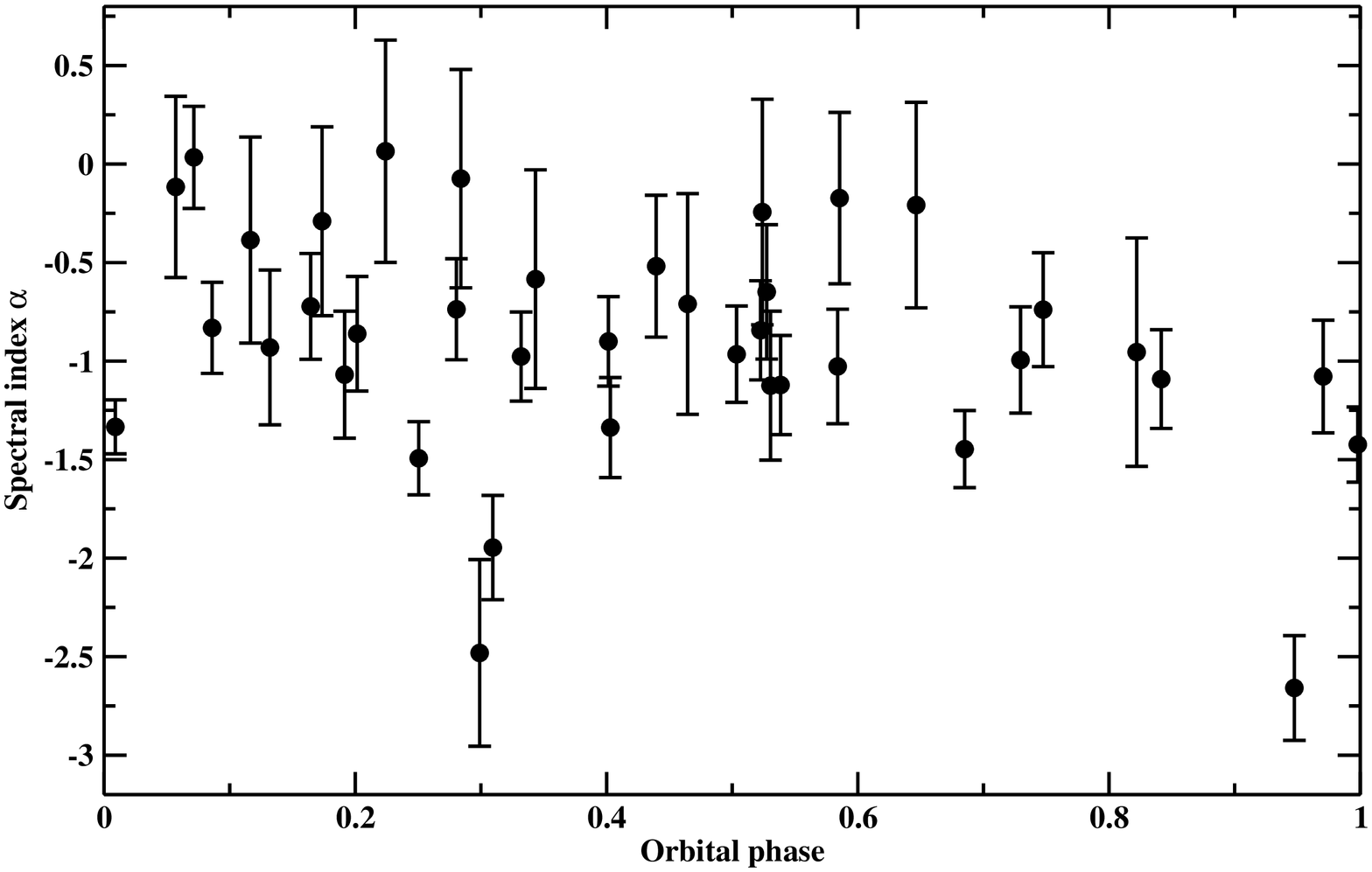}
  \caption{The global spectral index of Circinus X-1 between 4.8 and 8.6 GHz as a function of the orbital phase (radio ephemeris from \citealt{Nic07}, 
during the period 1996-2006. The mean value and standard deviation of the spectral index are $\alpha=-0.9 \pm 0.6$ $(F_{\nu} \propto \nu^{\alpha})$.}
\end{figure}

\section{Polarization properties}

Polarization was detected at more than 3 $\sigma$ confidence level on 13 epochs at 4.8 GHz and 1 epoch at 8.6 GHz. Faraday rotation (i.e. rotation 
of the plane of polarization due to the propagation of radiation through magnetized plasma) was observed on epoch 21 at a level of -175 $\pm$ 30 
rad m$^{-1}$ (Fig. 11). A less confident detection ($<$ 3 $\sigma$), consistent with the previous one, was made on epoch 22. We warn the reader that 
N$\pi$ ambiguities might be affecting the calculations, especially since only 2 frequencies were available. The compact source 
J1520.6-571 located 7' away from Circinus X-1 (see section 5) is also polarized. Its Faraday rotation measure is +185 $\pm$ 30 rad m$^{-1}$. 
Since these two objects are so close on the plane of the sky, although the exact galactic or extragalactic nature of J1520.6-571 is not known, it is 
reasonable to assume that the galactic contribution to the detected Faraday rotation is about the same in both cases. Then, given the opposite sign 
of the rotation measure values observed in the two sources, the conclusion is that the galactic contribution has to be small in absolute value, of the 
order of a few tens of rad m$^{-1}$. Otherwise the intrinsic Faraday rotation measure (i.e. due to the immediate environment of the objects) would have to 
be fairly large for one or the other of the sources. Another hint suggesting a low galactic Faraday rotation measure towards Circinus X-1 
comes from the attempts to mapping the whole sky. Containing around 800 reliable Faraday rotation measure, the map of \cite*{Joh04} seems to favour a small 
in absolute value, likely negative rotation measure towards Circinus X-1 (galactic coordinates: l=322\fdg12; b=+0\fdg04). We can therefore 
tentatively estimate that the galactic Faraday rotation measure in the direction of Circinus X-1 is probably somewhere between -50 and 50 rad m$^{-1}$. 
If this is true, 
it means that most of the rotation measure in Fig. 11 is likely intrinsic, originating in a Faraday screen in the vicinity of the object. 

The polarization maps are shown in Fig. 12. Assuming a -50 / +50 rad m$^{-1}$ Faraday rotation measure within the galaxy, the electric vector position 
angles in Fig. 12 have to be corrected by rotating them counter-clockwise / clockwise by $\sim$11\degr at 4.8 GHz and $\sim$3\fdg5 at 8.6 GHz. For an optically 
thin emitting region (as it is the case; see section 6) the magnetic 
field vectors are practically perpendicular to the electric field vectors. Although it is difficult to make strong statements, in the jet the electric field 
vectors seem to be oriented preferentially along its 
axis (e.g. maps C - 12, 17, 22, 23 in Fig. 12) thus suggesting the presence of shocks at the interface between the ejected matter and the surrounding 
environment, while close to the core the electric vectors tend to be oriented perpendicular to the jet axis. Moreover, the orientation of the electric vectors 
position angles seems to be stable over the almost 10 year period covered by the observations. The fractional linear polarization is of the order of a few percents.
 
\begin{figure}
\begin{center}
  \includegraphics[angle=-90,scale=0.3]{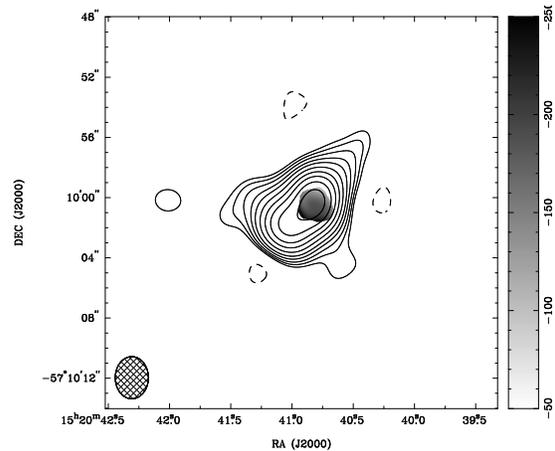}
  \caption{Faraday rotation map of Circinus X-1 from epoch 21 (2000 October 14/15). The contour lines correspond to the radio image at 4.8 GHz and are at 
-2.8, 2.8, 4, 5.6, 8, 11, 16, 23, 32, 45, 64, 90 $\times$ the rms noise. The size of the restoring beam is 2.8 $\times$ 2.2 arcsec$^2$, PA=0\fdg0. The 
gray code bar is expressed in rad m$^{-2}$. Only the region with a confident detection is shown. Please note that only two frequencies were used to generate 
the map.}
\end{center}
\end{figure}

\begin{figure*}
  \includegraphics[scale=0.23]{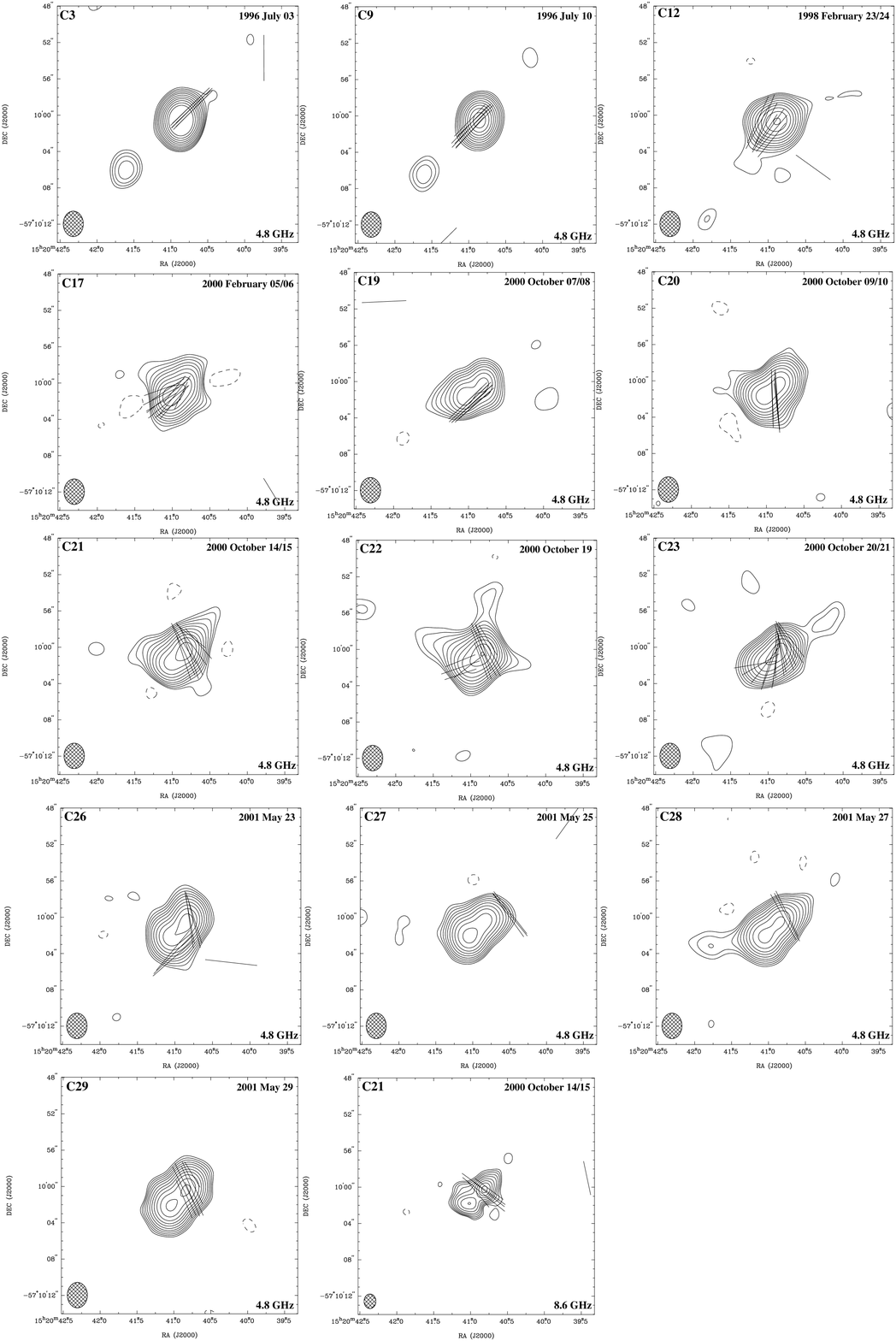}
  \caption{Radio polarization maps of Circinus X-1. The contour lines are at -2.8, 2.8, 4, 5.6, 8, 11, 16, 23, 32, 45, 64, 90 $\times$ the rms noise at 
each epoch. The size of the restoring beam is 2.8 $\times$ 2.2 arcsec$^2$, PA=0\fdg0 at 4.8 GHz and 1.6 $\times$ 1.3 arcsec$^2$, PA=0\fdg0 at 8.6 GHz. 
The sticks correspond to the orientation of the electric vector position angles. No correction for the Faraday rotation within our galaxy has been applied. 
For an optically thin emitting region the magnetic 
field vectors are practically perpendicular to the electric field vectors.}
\end{figure*}

\section{Conclusions}

We have presented ATCA radio data of the XRB Circinus X-1 at 4.8 and 8.6 GHz covering a period of time of almost 10 years. The radio maps reveal a complex 
structure of the radio emission with variations in the morphology and brightness of the different compact emitting regions (sometimes seen as far as 
10 arcsec from the core, corresponding to a physical scale of 0.05 $r$ pc, with $r$ the distance to Circinus X-1 expressed in kpc) at timescales of days. 
The orientation of the radio emission is on a NW-SE direction in the plane of the sky and remains 
unchanged during the period covered by our observations. This suggests that if the jet is indeed precessing, as proposed by \cite{Iar08}, then its 
precession period should be much larger then the time span of our data set. 

We were able to confirm the claim of \cite{Fen97} for a radio flare 
not associated with the orbital phase 0.0. Further evidence was found that such outbursts do occur near the orbital phase 0.5 (within a region of 
$\pm$ 0.1 in phase). Their properties are similar to those of the outbursts commonly observed near the orbital phase 0.0. We suggested that the 
wind accretion from the massive companion might be responsible for producing them. 

\cite{Fen04} showed that the outflow in the arcsec scale jet of Circinus X-1 is relativistic. Reanalyzing the data in some more detail and determining 
the time delay between the successive brightening of the various compact radio emitting regions, it was found that indeed the outflow seems to be relativistic 
with an apparent velocity $\beta_{app} \geq$ 12.

A study of the proper motion of the system was carried out. The errors in determining the positions of the core (dominated by the errors associated with 
the coordinates of the phase-referencing calibrator) allowed only an upper limit to be estimated, namely 50 mas yr$^{-1}$($v_{system} <$ 240 $r$ km 
s$^{-1}$, where $r$ is the distance to Circinus X-1 expressed in kpc). This is however a 
conservative value and future observations together with improvements in the systematic errors of the calibrators will pull down the limit and offer a better 
constraint. 

The global spectral index of Circinus X-1 (between 4.8 and 8.6 GHz) had a mean value and standard deviation of $\alpha=-0.9 \pm 0.6$ $(F_{\nu} \propto 
\nu^{\alpha})$ during the monitoring period. The spectrum was flattening when flares were active, either close to phase 0.0 or 0.5. 

Polarization emission was detected in one third of the epochs and in one epoch Faraday rotation was observed at a level of -175 $\pm$ 30 rad m$^{-1}$. The 
nature of the Faraday screen is unknown. Circumstantial evidence points towards an intrinsic origin, related to the immediate vicinity of Circinus X-1 
itself, however this has to be tested in detail in future observations. In the jet the electric field vectors seem to be oriented preferentially along its 
axis tentatively suggesting the presence of shocks at the interface between the ejected matter and the surrounding environment. 

\section*{Acknowledgments}

The Australia Telescope is funded by the Commonwealth of Australia for operation as a national facility managed by CSIRO. The X-ray data was 
provided by the ASM/RXTE teams at MIT and at the RXTE SOF and GOF at NASA's GSFC.

\appendix

\section{Image-plane fits}

The methodology employed was the following. We first fitted the data iteratively with a single Gaussian component and at each iteration subtracted the 
resulting model from the residual map obtained in the previous iteration. 
We then used the fitting parameters thus obtained as initial estimations in re-fitting the data this time with all the components simultaneously included. 
Tables A1 and A2 show the results obtained by fitting the natural weighted radio maps at 4.8 and 8.6 GHz respectively. 

For coordinates, the quoted errors take into account the systematic error associated with the position of the phase-referencing calibrator (250 mas)
and the formal errors due to the fitting process.

\begin{table*}
 \centering
  \caption{Results of the image-plane fitting at 4.8 GHz. The table contains the number of the epoch of observation, the corresponding date, the position 
(J2000) of the component fitted, its flux density (integrated flux density for Gaussian components or peak flux density for unresolved Gaussian 
components), size (major $\times$ minor axis), and orientation. The first entry for each epoch corresponds to the core of the system. For unresolved Gaussian 
components upper limits for the size are presented. The errors are reported in 
parentheses. For coordinates, the quoted errors take into account the systematic error associated with the position of the phase-referencing calibrator 
and the formal errors within the fitting process.}
   \begin{tabular}{@{}ccccccc}
  \hline \hline
   Ep  & Date & RA [h:m:s] & DEC [$\degr:\arcmin:\arcsec$] & $F_{\nu}$ [mJy] & Size [arcsec $\times$ arcsec] & PA 
[$\degr$] \\
\hline 
1 & 96/07/01 & 15:20:40.856 (0.032) & -57:10:00.498 (0.267) & 29.6 (0.4) & $<$ 2.8 $\times$ 2.2 & - \\
  &          & 15:20:41.141 (0.082) & -57:10:02.513 (0.720) & 1.1 (0.3) & $<$ 2.8 $\times$ 2.2 & - \\
  &          & 15:20:40.645 (0.093) & -57:09:56.844 (0.924) & 0.7 (0.3) & $<$ 2.8 $\times$ 2.2 & - \\ 
2 & 96/07/02 & 15:20:40.861 (0.032) & -57:10:00.545 (0.266) & 29.7 (0.3) & $<$ 2.8 $\times$ 2.2 & - \\
  &          & 15:20:40.446 (0.071) & -57:09:56.797 (0.639) & 1.6 (0.3) & $<$ 2.8 $\times$ 2.2 & - \\
  &          & 15:20:41.570 (0.072) & -57:10:05.709 (0.670) & 1.0 (0.2) & $<$ 2.8 $\times$ 2.2 & - \\
3 & 96/07/03 & 15:20:40.862 (0.031) & -57:10:00.582 (0.258) & 22.5 (0.1) & $<$ 2.8 $\times$ 2.2 & - \\
  &          & 15:20:41.610 (0.048) & -57:10:05.966 (0.416) & 1.0 (0.1) & 1.7 (0.4) $\times$ 0.7 (0.3) & -46.0 (21.6) \\
4 & 96/07/04 & 15:20:40.816 (0.032) & -57:10:00.271 (0.261) & 22.4 (0.2) & $<$ 2.8 $\times$ 2.2 & - \\
  &          & 15:20:39.679 (0.046) & -57:09:50.584 (0.385) & 2.0 (0.2) & $<$ 2.8 $\times$ 2.2 & - \\
5 & 96/07/05 & 15:20:40.865 (0.032) & -57:10:00.544 (0.267) & 13.8 (0.2) & $<$ 2.8 $\times$ 2.2 & - \\
  &          & 15:20:41.666 (0.051) & -57:10:05.440 (0.477) & 1.7 (0.1) & 2.8 (0.6) $\times$ 1.2 (0.4) & -25.9 (13.6) \\
6 & 96/07/07 & 15:20:40.902 (0.034) & -57:10:00.576 (0.282) & 6.2 (0.1) & 0.9 (0.1) $\times$ 0.4 (0.1) & -37.8 (4.4) \\
7 & 96/07/08 & 15:20:40.926 (0.032) & -57:10:00.862 (0.266) & 6.9 (0.1) & $<$ 2.8 $\times$ 2.2 & - \\
  &          & 15:20:41.666 (0.059) & -57:10:04.671 (0.492) & 0.5 (0.1) & $<$ 2.8 $\times$ 2.2 & - \\
8 & 96/07/09 & 15:20:40.896 (0.034) & -57:10:00.685 (0.280) & 9.6 (0.2) & 1.7 (0.1) $\times$ 0.5 (0.1) & -54.2 (4.5) \\
9 & 96/07/10 & 15:20:40.854 (0.032) & -57:10:00.483 (0.261) & 20.7 (0.2) & 0.8 (0.1) $\times$ 0.3 (0.1) & -50.4 (1.6) \\
  &          & 15:20:41.604 (0.050) & -57:10:06.415 (0.425) & 1.9 (0.1) & 2.6 (0.4) $\times$ 1.1 (0.4) & -47.5 (16.0) \\
10 & 96/07/13 & 15:20:41.065 (0.035) & -57:10:01.022 (0.292) & 8.5 (0.3) & 0.9 (0.1) $\times$ 0.6 (0.1) & -81.2 (7.6) \\
11 & 98/02/05-06 & 15:20:40.892 (0.036) & -57:10:00.560 (0.295) & 9.5 (0.2) & 2.1 (0.1) $\times$ 1.4 (0.1) & 88.4 (53.4) \\
12 & 98/02/23-24 & 15:20:40.894 (0.034) & -57:10:00.721 (0.276) & 10.5 (0.1) & 2.3 (0.1) $\times$ 0.8 (0.1) & -73.7 (5.6) \\
   &             & 15:20:41.803 (0.091) & -57:10:11.592 (0.894) & 0.3 (0.1) & $<$ 2.8 $\times$ 2.2 & - \\
13 & 98/10/03-04 & 15:20:40.892 (0.037) & -57:10:00.725 (0.289) & 7.9 (0.1) & 3.5 (0.1) $\times$ 0.8 (0.1) & -77.5 (3.3) \\
14 & 98/10/16-17 & 15:20:40.910 (0.037) & -57:10:00.700 (0.294) & 6.4 (0.1) & 3.1 (0.1) $\times$ 0.8 (0.1) & -75.3 (4.9) \\
15 & 98/10/29-30 & 15:20:40.886 (0.035) & -57:10:00.663 (0.283) & 8.2 (0.1) & 2.6 (0.1) $\times$ 0.7 (0.1) & -76.6 (5.9) \\
16 & 98/11/02 & 15:20:40.876 (0.037) & -57:10:00.663 (0.288) & 5.4 (0.1) & 3.1 (0.1) $\times$ 1.2 (0.1) & -87.9 (5.2) \\
17 & 00/02/05-06 & 15:20:40.791 (0.045) & -57:09:59.941 (0.381) & 4.3 (0.6) & $<$ 2.8 $\times$ 2.2 & - \\
   &             & 15:20:41.018 (0.040) & -57:10:01.496 (0.317) & 11.6 (0.3) & 1.6 (0.1) $\times$ 1.5 (0.1) & 13.5 (10.4) \\
18 & 00/10/01 & 15:20:40.739 (0.068) & -57:10:00.096 (0.437) & 5.6 (1.1) & $<$ 2.8 $\times$ 2.2 & - \\
   &          & 15:20:41.013 (0.058) & -57:10:01.692 (0.448) & 13.4 (0.6) & 2.5 (0.4) $\times$ 1.8 (0.2) & -73.8 (41.1) \\
19 & 00/10/07-08 & 15:20:40.768 (0.044) & -57:10:00.048 (0.409) & 5.6 (0.8) & $<$ 2.8 $\times$ 2.2 & - \\
   &             & 15:20:41.043 (0.050) & -57:10:01.578 (0.342) & 12.5 (0.4) & 2.4 (0.1) $\times$ 0.8 (0.3) & -74.8 (13.8) \\
20 & 00/10/09-10 & 15:20:40.828 (0.044) & -57:10:00.305 (0.361) & 10.8 (0.3) & 2.0 (0.2) $\times$ 0.6 (0.1) & -54.4 (6.4) \\
   &             & 15:20:41.068 (0.043) & -57:10:01.905 (0.310) & 9.3 (0.6) & $<$ 2.8 $\times$ 2.2 & - \\
21 & 00/10/14-15 & 15:20:40.835 (0.041) & -57:10:00.438 (0.339) & 19.9 (0.4) & 2.4 (0.2) $\times$ 0.9 (0.1) & -44.7 (4.5) \\
   &             & 15:20:41.102 (0.054) & -57:10:01.852 (0.355) & 6.6 (0.8) & $<$ 2.8 $\times$ 2.2 & - \\
22 & 00/10/19 & 15:20:40.880 (0.042) & -57:10:00.542 (0.341) & 23.5 (0.6) & 1.8 (0.2) $\times$ 1.2 (0.1) & -47.6 (9.1) \\
   &          & 15:20:41.104 (0.060) & -57:10:01.720 (0.417) & 9.0 (1.2) & $<$ 2.8 $\times$ 2.2 & - \\
23 & 00/10/20-21 & 15:20:40.812 (0.042) & -57:10:00.038 (0.460) & 11.6 (2.6) & $<$ 2.8 $\times$ 2.2 & - \\
   &             & 15:20:40.998 (0.066) & -57:10:01.650 (0.456) & 18.1 (1.6) & $<$ 2.8 $\times$ 2.2 & - \\
   &             & 15:20:40.243 (0.137) & -57:09:57.376 (0.855) & 3.1 (0.2) & 6.0 (2.7) $\times$ 1.9 (0.9) & -59.7 (16.9) \\
24 & 00/10/23 & 15:20:40.797 (0.046) & -57:10:00.087 (0.398) & 10.4 (0.7) & $<$ 2.8 $\times$ 2.2 & - \\
   &          & 15:20:41.032 (0.048) & -57:10:01.726 (0.341) & 14.5 (0.6) & 2.5 (0.1) $\times$ 0.2 (0.2) & 75.2 (10.1) \\
25 & 00/10/25-26 & 15:20:40.863 (0.061) & -57:10:00.619 (0.479) & 15.6 (0.8) & $<$ 2.8 $\times$ 2.2 & - \\
   &             & 15:20:41.088 (0.056) & -57:10:01.988 (0.391) & 9.8 (2.0) & $<$ 2.8 $\times$ 2.2 & - \\
   &             & 15:20:40.365 (0.092) & -57:09:57.374 (0.867) & 1.0 (0.4) & $<$ 2.8 $\times$ 2.2 & - \\
26 & 01/05/23 & 15:20:40.838 (0.039) & -57:10:00.443 (0.349) & 11.6 (0.3) & 1.8 (0.2) $\times$ 0.8 (0.1) & -24.3 (5.2) \\
   &          & 15:20:41.075 (0.045) & -57:10:02.322 (0.330) & 7.9 (0.5) & 1.8 (0.1) $\times$ 0.6 (0.2) & 51.1 (8.3) \\
27 & 01/05/25 & 15:20:40.816 (0.049) & -57:10:00.283 (0.378) & 6.3 (0.3) & $<$ 2.8 $\times$ 2.2 & - \\
   &          & 15:20:41.066 (0.041) & -57:10:02.138 (0.335) & 8.7 (0.4) & 1.7 (0.1) $\times$ 0.7 (0.1) & 71.5 (14.3) \\
28 & 01/05/27 & 15:20:40.842 (0.043) & -57:10:00.493 (0.348) & 9.2 (0.2) & $<$ 2.8 $\times$ 2.2 & - \\
   &          & 15:20:41.085 (0.040) & -57:10:02.380 (0.319) & 7.0 (0.5) & $<$ 2.8 $\times$ 2.2 & - \\
   &          & 15:20:41.692 (0.153) & -57:10:03.020 (0.478) & 1.6 (0.1) & $<$ 2.8 $\times$ 2.2 & - \\
29 & 01/05/29 & 15:20:40.814 (0.036) & -57:10:00.214 (0.304) & 7.7 (0.1) & 1.4 (0.1) $\times$ 0.7 (0.1) & -46.3 (4.9) \\
   &          & 15:20:41.065 (0.037) & -57:10:02.234 (0.293) & 8.4 (0.2) & 1.5 (0.1) $\times$ 1.2 (0.1) & 78.2 (8.0) \\
30 & 01/09/08 & 15:20:40.833 (0.058) & -57:10:00.105 (0.509) & 5.0 (0.2) & 3.0 (0.3) $\times$ 1.5 (0.4) & 31.7 (11.9) \\
   &          & 15:20:41.056 (0.054) & -57:10:02.351 (0.450) & 4.5 (0.3) & 3.0 (0.3) $\times$ 0.6 (0.4) & 41.4 (8.1) \\
\hline
\end{tabular}
\end{table*}

\begin{table*}
 \centering
  \contcaption{}
   \begin{tabular}{@{}ccccccc}
  \hline 
   Ep  & Date & RA [h:m:s] & DEC [$\degr:\arcmin:\arcsec$] & $F_{\nu}$ [mJy] & Size [arcsec $\times$ arcsec] & PA 
[$\degr$] \\
\hline
31 & 02/12/02-03 & 15:20:40.749 (0.050) & -57:10:00.119 (0.450) & 1.4 (0.2) & $<$ 2.8 $\times$ 2.2 & - \\
   &             & 15:20:41.060 (0.054) & -57:10:02.451 (0.487) & 1.1 (0.2) & $<$ 2.8 $\times$ 2.2 & - \\ 
32 & 02/12/03-04 & 15:20:40.749 (0.051) & -57:10:00.270 (0.452) & 1.4 (0.2) & $<$ 2.8 $\times$ 2.2 & - \\
   &             & 15:20:41.009 (0.060) & -57:10:02.129 (0.519) & 1.1 (0.2) & $<$ 2.8 $\times$ 2.2 & - \\
   &             & 15:20:41.309 (0.066) & -57:10:00.277 (0.592) & 0.8 (0.2) & $<$ 2.8 $\times$ 2.2 & - \\ 
33 & 02/12/04-05 & 15:20:40.842 (0.038) & -57:10:00.670 (0.316) & 11.1 (0.3) & $<$ 2.8 $\times$ 2.2 & - \\
   &             & 15:20:41.447 (0.082) & -57:10:00.152 (0.770) & 1.0 (0.3) & $<$ 2.8 $\times$ 2.2 & - \\
   &             & 15:20:40.313 (0.073) & -57:10:02.107 (0.667) & 1.2 (0.3) & $<$ 2.8 $\times$ 2.2 & - \\
34 & 02/12/05-06 & 15:20:40.839 (0.034) & -57:10:00.518 (0.292) & 10.6 (0.2) & $<$ 2.8 $\times$ 2.2 & - \\
   &             & 15:20:41.220 (0.067) & -57:10:03.509 (0.609) & 0.9 (0.2) & $<$ 2.8 $\times$ 2.2 & - \\
35 & 02/12/06-07 & 15:20:40.834 (0.036) & -57:10:00.446 (0.321) & 5.4 (0.1) & $<$ 2.8 $\times$ 2.2 & - \\
   &             & 15:20:41.129 (0.069) & -57:10:03.140 (0.444) & 1.5 (0.1) & 3.2 (0.5) $\times$ 1.2 (0.5) & 70.3 (21.02) \\
   &             & 15:20:41.796 (0.071) & -57:10:10.397 (0.667) & 0.35 (0.09) & $<$ 2.8 $\times$ 2.2 & - \\
36 & 02/12/07-08 & 15:20:40.866 (0.045) & -57:10:00.793 (0.369) & 4.1 (0.1) & 4.7 (0.4) $\times$ 1.5 (0.2) & -46.0 (4.1) \\
   &             & 15:20:41.388 (0.075) & -57:10:02.634 (0.650) & 0.5 (0.1) & $<$ 2.8 $\times$ 2.2 & - \\
37 & 02/12/08-09 & 15:20:40.775 (0.092) & -57:09:59.857 (0.952) & 0.5 (0.2) & $<$ 2.8 $\times$ 2.2 & - \\
   &             & 15:20:40.936 (0.075) & -57:10:01.364 (0.633) & 3.1 (0.1) & 6.6 (0.9) $\times$ 1.7 (0.4) & -45.3 (6.5) \\
38 & 03/12/23-24 & 15:20:40.822 (0.071) & -57:10:00.495 (0.671) & 0.9 (0.2) & $<$ 2.8 $\times$ 2.2 & - \\
   &             & 15:20:41.075 (0.074) & -57:10:03.070 (0.692) & 0.9 (0.2) & $<$ 2.8 $\times$ 2.2 & - \\
39 & 05/04/06 & 15:20:40.780 (0.063) & -57:10:00.737 (0.579) & 0.8 (0.2) & $<$ 2.8 $\times$ 2.2 & - \\
   &          & 15:20:41.784 (0.083) & -57:10:11.419 (0.791) & 0.47 (0.15) & $<$ 2.8 $\times$ 2.2 & - \\
40 & 05/06/17 & 15:20:40.841 (0.032) & -57:10:00.561 (0.265) & 38.8 (0.5) & $<$ 2.8 $\times$ 2.2 & - \\
   &             & 15:20:40.453 (0.065) & -57:10:03.678 (0.607) & 1.7 (0.3) & $<$ 2.8 $\times$ 2.2 & - \\
   &             & 15:20:41.130 (0.081) & -57:09:56.866 (0.779) & 1.1 (0.3) & $<$ 2.8 $\times$ 2.2 & - \\
41 & 06/03/22-23 & - & - & - & - & - \\
\hline \hline
\end{tabular}
\end{table*}

\begin{table*}
 \centering
  \caption{Results of the image-plane fitting at 8.6 GHz.  The table contains the number of the epoch of observation, the corresponding date, the position 
(J2000) of the component fitted, its flux density (integrated flux density for Gaussian components or peak flux density for unresolved Gaussian 
components), size (major $\times$ minor axis), and orientation. The first entry for each epoch corresponds to the core of the system. For unresolved Gaussian 
components upper limits for the size are presented. The errors are reported in 
parentheses. For coordinates, the quoted errors take into account the systematic error associated with the position of the phase-referencing calibrator 
and the formal errors within the fitting process.}
   \begin{tabular}{@{}ccccccc}
  \hline \hline
   Ep  & Date & RA [h:m:s] & DEC [$\degr:\arcmin:\arcsec$] & $F_{\nu}$ [mJy] & Size [arcsec $\times$ arcsec] & PA 
[$\degr$] \\
\hline
1 & 96/07/01 & 15:20:40.847 (0.033) & -57:10:00.483 (0.264) & 22.1 (0.4) & 0.4 (0.1) $\times$ 0.2 (0.1) & -8.3 (3.7) \\
  &          & 15:20:40.983 (0.058) & -57:10:00.910 (0.406) & 1.9 (0.7) & $<$ 1.6 $\times$ 1.3 & - \\ 
2 & 96/07/02 & 15:20:40.854 (0.032) & -57:10:00.537 (0.262) & 24.5 (0.4) & 0.6 (0.1) $\times$ 0.2 (0.1) & -47.8 (3.6) \\
  &          & 15:20:41.256 (0.054) & -57:10:03.360 (0.488) & 1.1 (0.3) & $<$ 1.6 $\times$ 1.3 & - \\
  &          & 15:20:40.601 (0.051) & -57:09:58.230 (0.472) & 1.5 (0.4) & $<$ 1.6 $\times$ 1.3 & - \\
3 & 96/07/03 & 15:20:40.851 (0.031) & -57:10:00.566 (0.256) & 23.6 (0.2) & $<$ 1.6 $\times$ 1.3 & - \\
  &          & 15:20:41.392 (0.047) & -57:10:04.468 (0.400) & 2.5 (0.1) & 3.1 (0.4) $\times$ 0.8 (0.2) & -43.0 (6.5) \\
4 & 96/07/04 & 15:20:40.817 (0.032) & -57:10:00.335 (0.260) & 20.0 (0.2) & $<$ 1.6 $\times$ 1.3 & - \\
  &          & 15:20:40.124 (0.042) & -57:09:54.482 (0.345) & 3.9 (0.2) & $<$ 1.6 $\times$ 1.3 & - \\
5 & 96/07/05 & 15:20:40.849 (0.032) & -57:10:00.502 (0.260) & 10.1 (0.1) & $<$ 1.6 $\times$ 1.3 & - \\
  &          & 15:20:41.441 (0.054) & -57:10:03.781 (0.412) & 1.4 (0.1) & 2.5 (0.5) $\times$ 0.9 (0.3) & -60.1 (13.8) \\
6 & 96/07/07 & 15:20:40.849 (0.034) & -57:10:00.267 (0.276) & 3.4 (0.1) & $<$ 1.6 $\times$ 1.3 & - \\
  &          & 15:20:40.973 (0.047) & -57:10:00.943 (0.380) & 0.7 (0.1) & $<$ 1.6 $\times$ 1.3 & - \\
7 & 96/07/08 & 15:20:40.885 (0.032) & -57:10:00.689 (0.262) & 5.3 (0.1) & 0.2 (0.1) $\times$ 0.1 (0.1) & 10.6 (3.5) \\
  &          & 15:20:41.102 (0.043) & -57:10:01.715 (0.378) & 0.7 (0.1) & 1.1 (0.3) $\times$ 0.3 (0.2) & 0.9 (18.1) \\
  &          & 15:20:40.501 (0.045) & -57:09:58.457 (0.377) & 0.5 (0.1) & $<$ 1.6 $\times$ 1.3 & - \\
8 & 96/07/09 & 15:20:40.860 (0.035) & -57:10:00.574 (0.276) & 6.5 (0.2) & 0.7 (0.1) $\times$ 0.2 (0.1) & -28.1 (5.5) \\
  &          & 15:20:41.022 (0.051) & -57:10:01.357 (0.379) & 1.7 (0.2) & 0.9 (0.2) $\times$ 0.4 (0.2) & -24.8 (21.7) \\
  &          & 15:20:40.639 (0.047) & -57:09:57.470 (0.413) & 0.5 (0.1) & $<$ 1.6 $\times$ 1.3 & - \\
9 & 96/07/10 & 15:20:40.848 (0.031) & -57:10:00.472 (0.255) & 17.3 (0.1) & 0.5 (0.1) $\times$ 0.1 (0.01) & -46.1 (1.4) \\
  &          & 15:20:41.288 (0.036) & -57:10:03.928 (0.298) & 1.9 (0.1) & 0.8 (0.1) $\times$ 0.3 (0.1) & -51.8 (14.4) \\
  &          & 15:20:40.471 (0.043) & -57:09:56.609 (0.377) & 0.7 (0.1) & 0.6 (0.3) $\times$ 0.3 (0.2) & -30.5 (33.0) \\
10 & 96/07/13 & 15:20:41.075 (0.033) & -57:10:01.068 (0.273) & 4.6 (0.1) & 0.5 (0.054) $\times$ 0.3 (0.1) & -65.0 (8.2) \\
   &          & 15:20:40.572 (0.045) & -57:09:59.757 (0.400) & 0.7 (0.1) & $<$ 1.6 $\times$ 1.3 & - \\
11 & 98/02/05-06 & 15:20:40.848 (0.033) & -57:10:00.421 (0.279) & 4.0 (0.1) & 0.7 (0.1) $\times$ 0.3 (0.1) & -15.9 (5.4) \\
   &             & 15:20:40.994 (0.058) & -57:10:00.697 (0.416) & 3.8 (0.1) & 5.0 (0.5) $\times$ 2.3 (0.3) & 44.1 (6.0) \\
12 & 98/02/23-24 & 15:20:40.851 (0.035) & -57:10:00.570 (0.274) & 5.7 (0.1) & 1.1 (0.1) $\times$ 0.4 (0.1) & -68.8 (8.8) \\
   &             & 15:20:41.046 (0.050) & -57:10:01.004 (0.361) & 1.7 (0.1) & 2.2 (0.3) $\times$ 0.9 (0.2) & 24.9 (8.4) \\
13 & 98/10/03-04 & 15:20:40.834 (0.051) & -57:10:00.455 (0.327) & 2.9 (0.1) & $<$ 1.6 $\times$ 1.3 & - \\
   &             & 15:20:41.082 (0.068) & -57:10:01.209 (0.357) & 1.2 (0.2) & 2.1 (0.3) $\times$ 0.3 (0.2) & 83.7 (15.3) \\
14 & 98/10/16-17 & 15:20:40.806 (0.041) & -57:10:00.304 (0.293) & 2.3 (0.1) & 1.7 (0.2) $\times$ 0.7 (0.1) & 89.0 (11.8) \\
   &             & 15:20:41.058 (0.040) & -57:10:00.929 (0.313) & 1.8 (0.1) & 1.8 (0.1) $\times$ 0.8 (0.1) & 40.0 (7.3) \\
15 & 98/10/29-30 & 15:20:40.831 (0.036) & -57:10:00.284 (0.329) & 1.6 (0.2) & $<$ 1.6 $\times$ 1.3 & - \\
   &             & 15:20:40.894 (0.045) & -57:10:00.801 (0.317) & 2.9 (0.1) & 3.2 (0.3) $\times$ 0.4 (0.1) & -78.9 (3.6) \\
16 & 98/11/02 & 15:20:40.783 (0.044) & -57:10:00.513 (0.313) & 2.0 (0.1) & 1.9 (0.2) $\times$ 0.6 (0.2) & 81.7 (9.8) \\
   &          & 15:20:40.993 (0.064) & -57:10:01.372 (0.482) & 1.8 (0.1) & 4.6 (0.7) $\times$ 1.7 (0.4) & 46.4 (6.7) \\
17 & 00/02/05-06 & 15:20:40.819 (0.035) & -57:10:00.147 (0.293) & 3.6 (0.1) & 1.2 (0.1) $\times$ 0.4 (0.1) & -32.8 (6.5) \\
   &             & 15:20:41.038 (0.034) & -57:10:01.637 (0.279) & 5.4 (0.1) & 1.1 (0.1) $\times$ 0.9 (0.1) & 14.0 (7.3) \\
18 & 00/10/01 & 15:20:40.779 (0.039) & -57:10:00.136 (0.314) & 4.2 (0.1) & 1.9 (0.2) $\times$ 0.9 (0.1) & 89.0 (12.5) \\
   &          & 15:20:41.014 (0.040) & -57:10:01.873 (0.298) & 5.5 (0.1) & 2.6 (0.2) $\times$ 0.8 (0.1) & 76.7 (4.6) \\
19 & 00/10/07-08 & 15:20:40.782 (0.043) & -57:10:00.155 (0.325) & 4.0 (0.2) & 2.2 (0.3) $\times$ 0.3 (0.1) & -63.5 (6.3) \\
   &             & 15:20:41.066 (0.038) & -57:10:01.795 (0.297) & 4.9 (0.2) & 1.4 (0.1) $\times$ 0.6 (0.1) & 80.8 (20.7) \\
20 & 00/10/09-10 & 15:20:40.806 (0.034) & -57:10:00.165 (0.283) & 4.8 (0.1) & 1.0 (0.1) $\times$ 0.9 (0.1) & 53.0 (11.6) \\
   &             & 15:20:41.038 (0.035) & -57:10:01.824 (0.278) & 6.2 (0.1) & 1.7 (0.1) $\times$ 1.0 (0.1) & 75.4 (10.0) \\
21 & 00/10/14-15 & 15:20:40.808 (0.034) & -57:10:00.187 (0.282) & 8.2 (0.2) & 1.4 (0.1) $\times$ 0.7 (0.1) & -54.4 (6.3) \\
   &             & 15:20:41.044 (0.037) & -57:10:01.789 (0.290) & 6.4 (0.2) & 1.8 (0.1) $\times$ 0.8 (0.1) & 62.2 (6.7) \\
22 & 00/10/19 & 15:20:40.835 (0.035) & -57:10:00.374 (0.289) & 12.5 (0.3) & 1.3 (0.1) $\times$ 0.9 (0.1) & 13.1 (7.7) \\
   &          & 15:20:41.021 (0.040) & -57:10:02.017 (0.339) & 3.5 (0.3) & $<$ 1.6 $\times$ 1.3 & - \\
   &          & 15:20:40.771 (0.051) & -57:09:57.183 (0.458) & 1.4 (0.3) & $<$ 1.6 $\times$ 1.3 & - \\
23 & 00/10/20-21 & 15:20:40.823 (0.035) & -57:10:00.320 (0.288) & 10.7 (0.3) & $<$ 1.6 $\times$ 1.3 & - \\
   &             & 15:20:41.037 (0.040) & -57:10:01.778 (0.300) & 7.5 (0.2) & $<$ 1.6 $\times$ 1.3 & - \\
   &             & 15:20:40.517 (0.047) & -57:09:58.979 (0.412) & 1.2 (0.2) & $<$ 1.6 $\times$ 1.3 & - \\
24 & 00/10/23 & 15:20:40.799 (0.035) & -57:10:00.165 (0.279) & 5.8 (0.1) & 1.5 (0.1) $\times$ 0.5 (0.1) & 68.7 (6.4) \\
   &          & 15:20:41.034 (0.035) & -57:10:01.772 (0.274) & 8.5 (0.1) & 2.2 (0.1) $\times$ 0.8 (0.1) & 68.0 (3.2) \\
25 & 00/10/25-26 & 15:20:40.804 (0.038) & -57:10:00.123 (0.311) & 5.4 (0.2) & 1.4 (0.1) $\times$ 0.8 (0.1) & 88.8 (51.2) \\
   &             & 15:20:41.058 (0.038) & -57:10:01.703 (0.293) & 8.9 (0.2) & 2.1 (0.1) $\times$ 1.0 (0.1) & -89.5 (8.5) \\
   &             & 15:20:40.524 (0.051) & -57:09:58.650 (0.449) & 0.9 (0.2) & $<$ 1.6 $\times$ 1.3 & - \\
\hline 
\end{tabular}
\end{table*}

\begin{table*}
 \centering
  \contcaption{}
   \begin{tabular}{@{}ccccccc}
  \hline 
   Ep  & Date & RA [h:m:s] & DEC [$\degr:\arcmin:\arcsec$] & $F_{\nu}$ [mJy] & Size [arcsec $\times$ arcsec] & PA 
[$\degr$] \\
\hline
26 & 01/05/23 & 15:20:40.836 (0.032) & -57:10:00.390 (0.272) & 7.2 (0.1) & 1.3 (0.1) $\times$ 0.3 (0.1) & -16.1 (2.2) \\
   &          & 15:20:41.058 (0.035) & -57:10:02.296 (0.279) & 5.5 (0.1) & 1.6 (0.1) $\times$ 0.9 (0.1) & 63.1 (7.0) \\ 
27 & 01/05/25 & 15:20:40.817 (0.035) & -57:10:00.315 (0.280) & 4.3 (0.1) & 1.5 (0.1) $\times$ 0.5 (0.1) & -85.3 (9.5) \\
   &          & 15:20:41.060 (0.034) & -57:10:02.102 (0.281) & 5.5 (0.1) & 1.6 (0.1) $\times$ 1.2 (0.1) & 65.4 (11.5) \\ 
28 & 01/05/27 & 15:20:40.818 (0.036) & -57:10:00.368 (0.292) & 4.6 (0.1) & 1.7 (0.1) $\times$ 0.3 (0.1) & -58.6 (4.7) \\
   &          & 15:20:41.063 (0.037) & -57:10:02.197 (0.292) & 5.5 (0.1) & 1.9 (0.1) $\times$ 1.3 (0.1) & 80.2 (15.5) \\
29 & 01/05/29 & 15:20:40.815 (0.034) & -57:10:00.209 (0.286) & 4.4 (0.1) & 1.5 (0.1) $\times$ 0.7 (0.1) & -21.2 (3.8) \\
   &          & 15:20:41.055 (0.035) & -57:10:02.164 (0.277) & 5.3 (0.1) & 1.6 (0.1) $\times$ 1.1 (0.1) & 80.0 (17.5) \\
30 & 01/09/08 & 15:20:40.844 (0.038) & -57:10:00.074 (0.327) & 2.1 (0.1) & 1.9 (0.2) $\times$ 0.4 (0.1) & 38.3 (6.5) \\
   &          & 15:20:41.065 (0.039) & -57:10:02.198 (0.317) & 2.7 (0.1) & 2.2 (0.2) $\times$ 0.7 (0.1) & 48.2 (5.6) \\
31 & 02/12/02-03 & 15:20:40.807 (0.045) & -57:10:00.058 (0.396) & 0.35 (0.05) & $<$ 1.6 $\times$ 1.3 & - \\
   &             & 15:20:41.010 (0.045) & -57:10:02.442 (0.397) & 0.34 (0.05) & $<$ 1.6 $\times$ 1.3 & - \\
32 & 02/12/03-04 & 15:20:40.819 (0.049) & -57:10:00.452 (0.392) & 1.7 (0.1) & 2.7 (0.4) $\times$ 2.1 (0.4) & -74.8 (43.8) \\
33 & 02/12/04-05 & 15:20:40.846 (0.034) & -57:10:00.619 (0.314) & 10.8 (0.5) & $<$ 1.6 $\times$ 1.3 & - \\
   &             & 15:20:41.095 (0.112) & -57:09:59.010 (0.590) & 3.7 (0.2) & 3.4 (1.4) $\times$ 1.2 (0.4) & 67.1 (15.4) \\
   &             & 15:20:40.587 (0.048) & -57:10:01.862 (0.419) & 1.5(0.3) & $<$ 1.6 $\times$ 1.3 & - \\
34 & 02/12/05-06 & 15:20:40.837 (0.032) & -57:10:00.496 (0.273) & 6.2 (0.1) & $<$ 1.6 $\times$ 1.3 & - \\
   &             & 15:20:41.117 (0.046) & -57:10:02.735 (0.408) & 0.6 (0.1) & $<$ 1.6 $\times$ 1.3 & - \\
35 & 02/12/06-07 & 15:20:40.835 (0.033) & -57:10:00.389 (0.283) & 2.7 (0.1) & $<$ 1.6 $\times$ 1.3 & - \\
   &             & 15:20:41.045 (0.061) & -57:10:02.769 (0.416) & 1.6 (0.1) & 4.1 (0.6) $\times$ 1.8 (0.3) & -75.5 (10.0) \\
   &             & 15:20:40.650 (0.052) & -57:09:59.398 (0.421) & 0.34 (0.07) & $<$ 1.6 $\times$ 1.3 & - \\
36 & 02/12/07-08 & 15:20:40.827 (0.041) & -57:10:00.455 (0.339) & 1.4 (0.1) & 2.5 (0.3) $\times$ 0.3 (0.1) & -47.8 (4.9) \\
   &             & 15:20:41.133 (0.045) & -57:10:03.096 (0.393) & 0.36 (0.05) & $<$ 1.6 $\times$ 1.3 & - \\
   &             & 15:20:40.691 (0.092) & -57:09:58.070 (0.966) & 0.09 (0.06) & $<$ 1.6 $\times$ 1.3 & - \\
37 & 02/12/08-09 & 15:20:40.846 (0.042) & -57:10:00.573 (0.364) & 0.44 (0.05) & $<$ 1.6 $\times$ 1.3 & - \\
   &             & 15:20:41.012 (0.047) & -57:10:02.392 (0.416) & 0.30 (0.05) & $<$ 1.6 $\times$ 1.3 & - \\
   &             & 15:20:40.644 (0.050) & -57:09:58.952 (0.440) & 0.26 (0.05) & $<$ 1.6 $\times$ 1.3 & - \\
38 & 03/12/23-24 & 15:20:40.849 (0.053) & -57:10:00.399 (0.489) & 0.24 (0.06) & $<$ 1.6 $\times$ 1.3 & - \\
   &             & 15:20:40.945 (0.057) & -57:10:02.492 (0.524) & 0.19 (0.05) & $<$ 1.6 $\times$ 1.3 & - \\
   &             & 15:20:40.722 (0.058) & -57:09:58.897 (0.526) & 0.20 (0.05) & $<$ 1.6 $\times$ 1.3 & - \\
39 & 05/04/06 & - & - & - & - & - \\
40 & 05/06/17 & 15:20:40.841 (0.032) & -57:10:00.550 (0.264) & 36.1 (0.6) & $<$ 1.6 $\times$ 1.3 & - \\
   &          & 15:20:40.959 (0.050) & -57:09:58.450 (0.449) & 2.4 (0.5) & $<$ 1.6 $\times$ 1.3 & - \\
   &          & 15:20:40.645 (0.051) & -57:10:02.649 (0.454) & 2.2 (0.5) & $<$ 1.6 $\times$ 1.3 & - \\
41 & 06/03/22-23 & - & - & - & - & - \\
\hline \hline
\end{tabular}
\end{table*}

\end{document}